\documentclass[a4paper,fleqn,usenatbib]{mnras}

\usepackage{newtxtext,newtxmath}
\usepackage[T1]{fontenc}
\usepackage{ae,aecompl}

\usepackage{natbib}
\usepackage{epsf}
\usepackage{color}
\usepackage{amsmath}
\usepackage{subfigure}
\usepackage{footnote}
\usepackage{graphicx}
\usepackage{tabularx}

\title[The Formation of Counter-Rotating S0 Galaxies]{The Formation of S0 Galaxies with Counter-Rotating Neutral and
  Molecular Hydrogen}
\author[R. Bassett et al.]{
Robert Bassett$^1$\thanks{E-mail: robert.bassett@uwa.edu.au (ICRAR)},
Kenji Bekki$^1$,
Luca Cortese$^1$,
Warrick Couch$^2$
\\
$^1$International Centre for Radio Astronomy Research,
University of Western Australia, 7 Fairway, Crawley, WA 6009,
Australia\\
$^{2}$Australian Astronomical Observatory, PO Box 970, North Ryde, NSW
1670, Australia}

\date{Accepted XXX. Received YYY; in original form ZZZ}

\pubyear{2016}

\begin{document}
\label{firstpage}
\pagerange{\pageref{firstpage}--\pageref{lastpage}}
\maketitle

\begin{abstract} 
The observation of counter rotation in galaxies (i.e. gas that rotates
in the opposite direction to the stellar component or two co-spatial
stellar populations with opposite rotation) is becoming more
commonplace with modern integral field spectroscopic surveys. In this paper we explore the emergence of
counter-rotation (both stellar and gaseous) in S0 galaxies from smoothed-particle hydrodynamics
simulations of 1/10 mass ratio minor mergers between a $\sim$10$^{10.8}$
M$_{\odot}$ disk galaxy with a bulge-to-total ration of 0.17 and a gas rich companion (gas-to-stellar mass
fraction of 5.0). These simulations include a self-consistent treatment of gas
dynamics, star formation, the production/destruction of H$_{2}$ and
dust, and the time evolution of the interstellar radiation field. We explore the effect of retrograde versus prograde obits, gas
and bulge mass fractions of the primary galaxy, and orbital parameters
of the companion. The key requirement for
producing counter rotation in stars or gas in a merger remnant is a retrograde primary, while the relative spin of the
companion affects only the radial extent of the accreted gas. We
also find that including a significant amount of gas in the primary
can prevent the emergence of counter-rotating gas, although accreted
stars retain counter-rotation. Bulge mass and orbit have a secondary
effect, generally influencing the final distribution of accreted stars
and gas within the framework outlined above. In addition to our
primary focus of counter-rotating components in galaxies, we also make some
predictions regarding the SFRs, H$_{2}$ distributions, and dust
in minor merger remnants.
\end{abstract}

\begin{keywords}
galaxies: interactions -- galaxies: kinematics and dynamics
\end{keywords}

%%%%%%%%%%%%%%%%%%%%%%%%%%%%%%%%%%%%%%%%%%%%%%%%%%

%%%%%%%%%%%%%%%%% BODY OF PAPER %%%%%%%%%%%%%%%%%%

\section{Introduction}\label{section:intro}

Prior to the 1980's the concept of counter-rotating matter in galaxies, galaxies
which contain multiple dynamical components (stars and/or
gas) rotating in different directions, was primarily theoretical
\citep{lyndenbell60,toomre82,binney87}. The first 
example of counter-rotation in an observation was shown by
\citet{galletta87} who observed an SB0 galaxy to have gas rotating in 
retrograde orbits relative to the stars. Observation of two counter-rotating stellar
components in E7/S0 galaxy NGC 4550 by \citet{rubin92} came a few years
later. In retrospect, it is not surprising that these discoveries were both made for S0
galaxies as subsequent studies have shown that the percentage of
S0 galaxies that exhibit counter-rotation is 20-40\%
\citep{bertola92,kuijken96,pizzella04,davis11} compared to
< 8\% for Sa-Sbc galaxies \citep{kannappan01}. For a more thorough
historical review on the topic of
multispin galaxies see \citet{rubin94} and \citet{corsini14}. 

Having been established as a concrete reality, counter-rotating
components of 
galaxies next required a viable formation mechanism. 
Another important development occuring contemporaneously to the
discovery of such systems was the
recognition of the importance of galaxy interactions in the emerging
hierarchical model of galaxy evolution \citep[][among
others]{gunn72,press74,white78,davis85}. A number of numerical
simulations have shown some galaxy mergers do indeed produce
counter-rotation in galaxies
\citep{balcells90,hernquist91,thakar97,bendo00,dimatteo07} and, as they are a cosmological
necessity, mergers represent the prime candidate for the origin of
counter-rotation. This option is also attractive in
that mergers have also been shown as a mechanism that can contribute
to the formtion of S0 galaxies \citep[e.g.][]{mihos95,bekki98,querejeta15}, which may explain the prevalence of
counter-rotation in intermediate morphological types. 

Among studies of the emergence of counter-rotation in simulations
only \citet{thakar97} and \citet{mapelli15} demonstrate counter-rotation in
minor merger remnants. These simulations are able to preserve the disk of the
central galaxy resulting in S0, rather than elliptical,
remnants. While the work of \citet{thakar97} is focused on the
emergence of counter-rotation, the authors tested only three minor merger
simulations with only the dark matter content of the accreted
satellite varying between simulations. These simulations are also dark
matter only, N-body simulations, thus an update using state-of-the-art
hydrodynamical simulations is warranted. The more recent work of
\citet{mapelli15} was focused on building gas rings in S0 galaxies
through mergers with the emergence of counter-rotation being a
secondary trait that is only briefly discussed. Thus, a large parameter space
is yet to be explored in regards to the primary factors required to
produce counter-rotating components through minor mergers.

In this work we perform a
series of minor merger simulations using our original chemodynamical
code for galaxy evolution \citep{bekki13}. Our code represents a major step
forward compared to the sticky particle simulations of
\citet{thakar97}, allowing us to track the complex evolution of
different stellar components in our galaxies as well as their
multiphase interstellar medium. Another improvement in this study when
compared to \citet{thakar97} and \citet{mapelli15} is that we perform
a large number of merger simulations with the specific focus of
understanding how the initial conditions affect the emergence of
counter-rotation. Thus, our study represents the first systematic
study of the formation of S0 galaxies hosting counter-rotating gas
and/or stars through minor
mergers.  Our suite of simulations allows us to investigate
the effects of the spin orientation of the merging galaxies relative
to their orbit, the gas fraction of the primary galaxy, bulge mass of
the central galaxy, as well as the
orbital parameters themselves. Simulations such as these can help to pin down
the statistics of how many galaxy mergers are likely to result in
kinematic misalignments easily identifiable by IFS
observations. Furthermore, for those simulations resulting in
co-rotation of gas and stars, we aim to identify observables that may
be strongly indicative of past galaxy interactions in the absence of
obvious peculiarities in photometry and/or kinematics. This paper is
laid out as follows: In Section \ref{section:model} we describe our
simulation code and the initial conditions of our merger simulations,
in Section \ref{section:mapping} we briefly describe our method of
producing two dimensional maps from out simulated data, in Section \ref{section:results} we describe the outcomes of our
mergers and possible observational signatures, in Section
\ref{section:discussion} we provide some discussion and comparison
with previous works, and in Section \ref{section:conclusions} we
briefly summarise our conclusions.

\section{The model}\label{section:model}

\subsection{A simulation code}

We employ the simulation code that has been recently 
developed in our previous works 
(\citealp{bekki13}, B13; \citealp{bekki14}, B14) to investigate the
structure and kinematics of gas, dust, and stars
in galaxy mergers.
Since the details of the code are already given in B13 and B14,
we describe them briefly here.
Gravitational
calculations can be done on GPUs whereas other calculations (e.g., gas
dynamics, dust evolution,
and hydrodynamics) are performed using CPUs, thus the code can be run on
GPU-based clusters.
The hydrodynamical part of the code is based on
the smoothed-particle hydrodynamics (SPH) method for following the time
evolution of gas dynamics in galaxies.

The code allows us to investigate 
gas dynamics, star formation from gas,
${\rm H_2}$ formation on dust grains,
formation of dust grains in the stellar winds of supernovae (SNe)
and asymptotic giant branch (AGB) stars, 
time evolution of interstellar radiation field (ISRF),
growth and destruction processes
of dust in the interstellar
medium (ISM),  
and ${\rm H_2}$ photo-dissociation due to far ultra-violet (FUV)
light in a self-consistent manner.
The new simulation code does not include 
the effects of feedback of active galactic nuclei (AGN)
on ISM and the growth of supermassive black holes (SMBHs) in galaxies.
Since our main purpose is not to investigate the AGN feedback effects
on  the formation of counter-rotating gas rings, we consider 
our adoption of the code for the investigation of gas and dust properties
in S0 galaxies is appropriate.

\subsection{Merger progenitor disk galaxy}

\subsubsection{Basic Description}

The smaller (`companion') and larger (`primary')
  galaxies in a minor merger are represented by disk galaxies
with the latter also including a central bulge component.
The total masses of dark matter halo, stellar disk, gas disk, and
bulge of a disk galaxy are denoted as $M_{\rm h}$, $M_{\rm s}$, $M_{\rm g}$,
and $M_{\rm b}$, respectively. The
gas mass fraction
is denoted as $f_{\rm g}$ and considered to be a key parameter that
determines the final kinematics of gas.
The evolution of gas in minor mergers
strongly depends on  $f_{\rm g}$ of the primary and the companion
($f_{\rm g, p}$ and $f_{\rm g, c}$, respectively),
and accordingly we investigate models with different $f_{\rm g, p}$
and $f_{\rm g, c}$.
We adopt the density distribution of the NFW
halo \citep*{navarro96} suggested from CDM simulations to describe the
initial density profile of dark matter halo
in a disk galaxy:
\begin{equation}
{\rho}(r)=\frac{\rho_{0}}{(r/r_{\rm s})(1+r/r_{\rm s})^2},
\end{equation}
Here,  $r$, $\rho_{0}$, and $r_{\rm s}$ are
the spherical radius,  the characteristic  density of a dark halo,  and the
scale
length of the halo, respectively.
The $c$-parameter ($c=r_{\rm vir}/r_{\rm s}$, where $r_{\rm vir}$ is the virial
radius of a dark matter halo) and $r_{\rm vir}$ are chosen appropriately
for a given dark halo mass ($M_{\rm dm}$)
by using recent predictions from cosmological simulations 
\citep[e.g.][]{neto07}.

The bulge of a disk galaxy 
is represented by the Hernquist
density profile with the bulge mass fraction ($M_{\rm b}/M_{\rm s}$)
as a free parameter ranging from 0 to 1. For the companion galaxy
in a minor merger, $f_{\rm b}$ is set to be 0 (bulge-less  disk
galaxy) while the primary galaxy, in most cases, contains a
non-rotating bulge with $f_{\rm b}$ = 0.17.
The radial ($R$) and vertical ($Z$) density profiles of the stellar disk are
assumed to be proportional to $\exp (-R/R_{0}) $ with scale
length $R_{0} = 0.2R_{\rm s}$  and to ${\rm sech}^2 (Z/Z_{0})$ with scale
length $Z_{0} = 0.04R_{\rm s}$, respectively.
The gas disk with a size  of $R_{\rm g}$
has the  radial and vertical scale lengths
of $0.2R_{\rm g}$ and $0.02R_{\rm g}$, respectively.
In the present model,  the exponential disk
has $R_{\rm s}=17.5$ kpc for the primary
and 5.5 kpc for the companion. The primary and companion galaxies
are assumed to have $R_{\rm g}/R_{\rm s}=1$ and 3, respectively.
In addition to the
rotational velocity caused by the gravitational field of the disk,
bulge, and dark halo components, the initial radial and azimuthal
velocity dispersions are assigned to the disc component according to
the epicyclic theory with Toomre's parameter $Q$ = 1.5.
The vertical velocity dispersion at a given radius is set to be 0.5
times as large as the radial velocity dispersion at that point.

In this work we describe the initial morphology of the primary galaxy
as being a late-type. In particular, these simulated galaxies
represent the early end of the late-type galaxy sequence (i.e. Sa
galaxies) for two reasons: first
they host large bulges and second the majority begin the simulations
free of gas. From an observational perspective, it may be equally valid to
describe our primary galaxies as initially having S0 morphologies
though this simply reflects the often ambiguous nature of visual
classifications for S0-Sa galaxies
\citep[e.g.][]{sandage04,kormendy12}. In simulations, the assertion
that a gas-free, disk-like galaxies should all be considered S0 in
morphology is not clear cut, with examples of simulated gas-free
galaxies hosting spiral arms readily available in the literature
\citep[e.g.][]{bekki02}. For this reason we choose to describe the
initial morphologies as late-type, although it can easily be argued
that an S0 classification is also appropriate. Regardless, we test
here the conditions required such that gas-rich, minor mergers
involving massive disks (whether Sa or S0) produce
S0 remnants containing counter-rotating gas and/or stars.

The initial gravitational softening lengths for dark matter halo, stellar disk, gas disk, and bulge
in the central galaxy are set to be 2.1 kpc, 0.2 kpc, 0.08 kpc, and 0.2 kpc, respectively.
Those for the dark matter halo, the stellar disk, and the gas disk of
the companion galaxy are set to be
0.88 kpc, 0.08 kpc, and 0.08 kpc, respectively. We employ a variable
softening length for dark matter and stellar particles residing in
dense regions, with a minimum value of 0.08 kpc, matching that of the
gas particles. Using these multiple and varying softening lengths
for different components, we are able to resolve the 100-1000pc scale
distribution of cold neutral and molecular gas in simulated S0s in
this present study. The main results of this work focus on the kpc
(and greater) scale kinematics of different components of our merger
remnants. Furthermore, as described in Section \ref{section:mapping},
we also employ a 2D, 1 kpc Gaussian smoothing kernal while producing
kinematics maps of our simulations. Thus, our conclusions based on large
scale kinematics should not change were we to rerun our simulations at
a much higher resolution (although such simulations may differ
significantly on sub-kpc scales). 

Other properties of the galaxies in our models are chosen to
roughly match those observed in the local universe. First, the initial
stellar mass of the primary galaxies of log$_{\mathrm{10}}$($M_{*}$) =
10.85 ($M_{\odot}$) is near the peak in stellar mass for S0 galaxies
observed in the GAMA survey \citep{moffett16}. As we mentioned
previously, we consider our primary galaxies to begin as late-type
galaxies, however we match their initial mass to those of low $z$ S0's
in order that the remnants will have masses appropriate for an S0 galaxy
(as there is little mass growth during the simulation). The corresponding mass
of the satellite galaxies give a merger mass ratio of 1/10, which
falls in the observed range for low redshift minor mergers
\citep{croton06,stewart08,somerville08,lotz11}. Finally, the gas mass
fractions of the companion galaxies of 0.5 are consistent with observed
values for dwarf galaxies from the ALFA survey presented in
\citet{huang12}. We note that the sample of \citet{huang12} will be
biased towards gas-rich systems, particularly at low stellar mass,
thus they likely represent the upper end in gas fraction compared to
the bulk population of dwarf galaxies. Regardless, gas-rich galaxies
similar to our simulated companions are known to exist and, although
likely rare, gas-rich minor mergers similar to our simulations will
occur and have been shown to be necessary to reproduce observations
\citep[e.g. the HI distributions of HI-excess galaxies, e.g.][]{gereb16}. 

Following the observed mass-metallicity relation in disk galaxies
\citep[e.g.][]{tremonti04}, we allocate a metallicity to each disk galaxy.
We do not assume a metallicity gradient in the present study in order 
to avoid introducing another model parameters that can hamper the 
interpretation of the simulation results.
Metallicity-dependent radiative cooling is included, and the initial
gas temperature is set to be 10,000 K for the primary and secondary galaxies
in the mergers. Gas-to-dust-ratio is a function of metallicity and we
assume a dust-to-metal
ratio of 0.4 for all models. The details of the model
for gas dynamics with dust is given in B13.

\subsubsection{Possible Limitations}

We are limited in the
exact types of mergers we can study due to the extremely large parameter space represented by the wide
variety of progenitor galaxies in minor mergers. First we point out that, although this work
represents a step forward towards a systematic analysis of the
emergence of counter-rotation in minor merger remnants, we primarily
focus on mergers involving gas poor (or gas free) primary galaxies with a fixed
bulge-to-total ratio of 0.17 and an extremely gas rich companion with
$f_{g,c}$ = 5.0. We focus on mergers such as these because the large gas
content of the merger remnants will be representative of those
galaxies hosting the most easily detectable counter-rotating gas
disks. While we do include few examples of mergers that vary these
basic parameters, we do not explore varying these values in a large
range of with a large range of initial conditions (e.g. low $f_{g,c}$
as well as $f_{b}$ = 1.0).

Recently, \citet{marinova12} showed that 50-65\% of S0 galaxies in the
local universe exhibit stellar bars. Although we find that bars form in a number
of our simulations (see Section \ref{section:morphs}), bars are not
included in the initial conditions for either the primary or companion
galaxy. Bars can significantly influence the angular momentum transfer
in galaxies, and thus may have an impact on the way material is
accreted during a minor merger. We expect any differences that may be
induced by an initial bar will be on sub-galactic scales and will not
affect the overall direction of rotation, and by extension our
classification of a given component as co- or counter-rotating. 

Finally, we emphasise that the bulge of the primary galaxy is
initially spherical and non-rotating. This is typical of bulges in hydrodynamical
simulations, but may not always be the case for observed galaxies that
can host flattened and/or rotating bulges \citep[e.g.][]{noordermeer08}. Although the
relative kinematics of primary and companion stars and gas can have an
effect on the final distribution \citet[e.g. enhanced tidal stripping in
prograde mergers][]{toomre72}, the fact that bulges in our simulations
are relatively small with $f_{b}$ = 0.17 means this can only have a
minor effect, if any, on our results. If anything, the presence of a
non-rotating component in the primary galaxy would act to confuse
signatures of counter-rotation by adding a random velocity component
to those stars of the primary galaxy disk. Thus, if the bulges in our
primary galaxies are initiated with rotation matching that of the
primary disk, the counter-rotation signature would only be enhanced. 

\subsection{Star formation}

We adopt the `${\rm H_2}$-dependent' SF recipe
(B13) in which SFR is determined by local molecular fraction
($f_{\rm H_2}$) for  each gas particle in the present study.
A gas particle {\it can be} converted
into a new star if the following three conditions are met
for each particle: (i) the local dynamical time scale is shorter
than the sound crossing time scale (mimicking
the Jeans instability) , (ii) the local velocity
field is identified as being consistent with gravitationally collapsing
(i.e., div {\bf v}$<0$),
and (iii) the local density exceeds a threshold density for star formation ($\rho_{\rm th}$).
We also adopt
the  Kennicutt-Schmidt law, which is described as 
SFR$\propto \rho_{\rm g}^{\alpha_{\rm sf}}$ \citep{kenni98},
where
$\alpha_{\rm sf}$ is the power-law slope.
A reasonable value of
$\alpha_{\rm sf}=1.5$ is adopted for all models.
The threshold gas density for star formation ($\rho_{\rm th}$) is 
set to be 1 cm$^{-3}$ for all models 
in the present study.

Each SN is assumed to eject the feedback energy ($E_{\rm sn}$)
of $10^{51}$ erg and 90\% and 10\% of $E_{\rm sn}$ are used for the increase
of thermal energy (`thermal feedback')
and random motion (`kinematic feedback'), respectively.
The thermal energy is used 
for the `adiabatic expansion phase', where each SN can remain
adiabatic for a timescale of $t_{\rm adi}$. This timescale is set to be
$10^6$ yr.
A canonical stellar initial mass function (IMF) proposed by \citet{kroupa01}, which
has three different slopes at different mass ranges is adopted.
and the IMF is assumed to be fixed at the canonical one.
Therefore, chemical evolution, SN feedback effects, and dust formation
and evolution is determined by the fixed IMF.

\subsection{Dust and metals}

Chemical enrichment through star formation and metal ejection from
SNIa, II, and AGB stars is self-consistently included in the chemodynamical
simulations.
The time evolution of the 11 chemical elements of H, He, C, N, O, Fe,
Mg, Ca, Si, S, and Ba is investigated to predict 
both chemical abundances and dust properties
in the present study, though we do not discuss these.
We consider the time delay between the epoch of star formation
and those  of supernova explosions and commencement of AGB phases (i.e.,
non-instantaneous recycling of chemical elements).
We adopt the nucleosynthesis yields of SNe II and Ia from 
\citet{tsujimoto95}
and AGB stars from \citet*{vandenhoek97}
in order to estimate chemical yields in the present study.

The dust model adopted in the present study is 
the same as those in B13 and B14:
The  total mass of $j$th component ($j$=C, O, Mg, Si, S, Ca, and Fe)
of dust from $k$th type of stars ($k$ = I, II, and AGB for SNe Ia, SNe II, and
AGB stars, respectively) are derived
based on the methods described in B13.
Dust can grow through accretion of existing metals onto dust grains
with a timescale of $\tau_{g}$.
Dust grains can be destroyed though supernova blast waves
in the ISM of galaxies
and the destruction process is parameterised by the destruction time scale
($\tau_{\rm d}$). We consider the models with
$\tau_{\rm g}=0.25$ Gyr and $\tau_{\rm d}=0.5$ Gyr, and the reason for
this selection is discussed in B13.

\subsection{${\rm H_2}$ formation and dissociation}

The details of the
new model for  ${\rm H_2}$ formation on dust grains
in galaxy-scale simulations have already been provided in B14,
therefore, we summarise only briefly here
the  model for ${\rm H_2}$ formation and dissociation in the present study.
The present chemodynamical simulations include both  ${\rm H_2}$ formation
on dust grains and ${\rm H}_2$ dissociation by FUV radiation
self-consistently.
The temperature ($T_{\rm g}$),
hydrogen density ($\rho_{\rm H}$),  dust-to-gas ratio ($D$)
of a gas particle and the strength of the
FUV radiation field ($\chi$) around the gas particle
are calculated at each time step so that the fraction of molecular
hydrogen ($f_{\rm H_2}$) for the gas particle can be derived based on
the ${\rm H_2}$ formation/destruction equilibrium conditions.
The SEDs of stellar particles around each $i$-th gas particles 
(thus ISRF) are first 
estimated from ages and metallicities of the stars by using stellar population
synthesis codes for a given IMF \citep[e.g.][]{bruzual03}.
Then the strength of the FUV-part of the ISRF
is estimated from the SEDs so that $\chi_i$ can be derived for the $i$-th gas particle.
Based on $\chi_i$, $D_i$, and  $\rho_{\rm H, \it i}$ of the gas particle,
we can derive $f_{\rm H_2, \it i}$ (See Fig. 1 in B13a).
Thus, each gas particle has $f_{\rm H_2, \it i}$, metallicity ([Fe/H]),
and gas density, and the total dust, metal, and 
 ${\rm H_2}$ masses are estimated
from these properties.

\subsection{Galaxy mergers}

%%%%% TABLE1
\begin{table}\label{table:1}
\centering
\begin{minipage}{\columnwidth}
\caption{Description of the basic parameter values
for the merger models.}
\begin{tabular}{|l|c|c|c|c|c|c|c|}
\hline
{ ID \footnote{ The isolated model is m1. 
 }} & 
{ $f_{\rm b}$ \footnote{ The initial bulge mass fraction of a primary galaxy.
 }} & 
{ $f_{\rm g,p}$ \footnote{ The initial gas mass ratio
($M_{\rm g}/M_{\rm s}$, where $M_{\rm g}$ and $M_{\rm s}$
are the initial total stellar and gas masses, respectively)  of a primary galaxy.
 }} & 
{ $f_{\rm g,c}$ \footnote{ The initial gas mass ratio of a companion galaxy.
 }} & 
{ Orbit \footnote{ `PP', `PR', `RP', and `RR' represent the 
prograde-prograde, prograde-retrograde, retrograde-prograde,
and retrograde-retrograde orbital configurations, respectively.
 }} & 
{ $r_{\rm p}$  \footnote{ The pericenter distance of a merger  
in units of kpc
}} &
{ $r_{\rm i}$  \footnote{ The initial distance of two disk galaxies
in units of kpc.
}} &
{ $e_{\rm o}$  \footnote{ The orbital eccentricity  of merging two galaxies.
}} \\
\hline
\hline
m1 & 0.17 & 0.09 & n/a & n/a & n/a & n/a & n/a\\
m2 & 0.17 & 0.0  & 5.0 & RR & 8.8 & 35.0 & 0.7 \\
m3 & 0.17 & 0.0  & 5.0 & PR & 8.8 & 35.0 & 0.7 \\
m4 & 0.17 & 0.0  & 5.0 & RP & 8.8 & 35.0 & 0.7 \\
m5 & 0.17 & 0.0  & 5.0 & PP & 8.8 & 35.0 & 0.7 \\
m6 & 0.17 & 0.01  & 5.0 & RR& 8.8 & 35.0 & 0.7 \\
m7 & 0.17 & 0.05  & 5.0 & RR& 8.8 & 35.0 & 0.7 \\
m8 & 0.17 & 0.1  & 5.0 & RR & 8.8 & 35.0 & 0.7 \\
m9 & 0.17 & 0.1  & 5.0 & PR& 8.8 & 35.0 & 0.7 \\
m10 & 0.17 & 0.0  & 5.0 & RR& 17.5 & 52.5 & 0.7 \\
m11 & 0.17 & 0.01  & 5.0 & RR& 17.5 & 52.5 & 0.7 \\
m12 & 0.17 & 0.05  & 5.0 & PP & 17.5 & 52.5 & 0.7 \\
m13 & 0.17 & 0.01  & 5.0 & RR& 17.5 & 175 & 1.0 \\
m14 & 1.0 & 0.01  & 5.0 & PR& 17.5 & 52.5 & 0.7 \\
m15 & 0.17 & 0.0  & 1.0 & RR & 8.8 & 35.0 & 0.7 \\
m16 & 0.17 & 0.0  & 0.5 & RR & 8.8 & 35.0 & 0.7 \\
\hline
\end{tabular}
\end{minipage}
\end{table}

\begin{figure}
  \centering
  \includegraphics[width=\columnwidth]{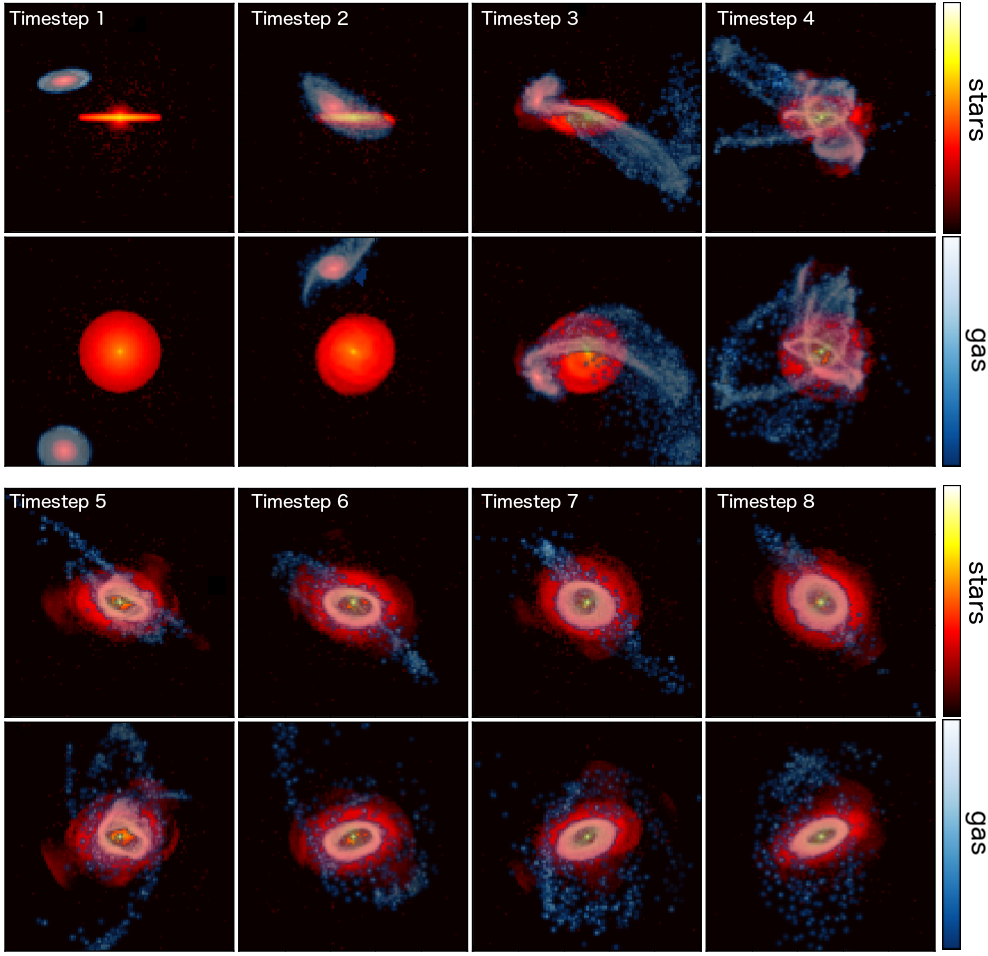}
  \caption{A qualitative picture of our general merger scenario in
    timesteps of 7.05$\times$10$^{8}$ yr for model m10. At each timestep we show two
  projections of the simulation stacked on top of eachother with a
  90$^{\circ}$ difference in viewing angle. In each panel we overplot
  gas and stars with separate colorbars that are indicated to the
  right.}\label{figure:simex}
\end{figure}

Both companion and the primary
galaxies  in a galaxy merger
are represented by the disk galaxy model
described above.
Although the mass-ratio of the companion to the primary can
 be  a free parameter represented
by $m_2$,
we investigate  only for the models with $m_2=0.1$.
This is because the simulated merger remnants can have 
S0-like morphology without spiral arms.

In all the simulations of minor mergers,
 the orbit of the two disks is set to be
initially on the $xy$ plane and the distance between
the center of mass of the two disks
is a free parameter ($R_{\rm i}$).
The pericenter
distance, represented by $r_{\rm p}$ 
and the orbital eccentricity ($e_{\rm o}$) 
are free parameters too.
The spin of the primary and companion  galaxies in a merger pair
is specified by two angle $\theta$ and
$\phi$ (in units of degrees), where
$\theta$ is the angle between the $z$ axis and the vector of
the angular momentum of a disk and
$\phi$ is the azimuthal angle measured from $x$ axis to
the projection of the angular momentum vector of a disk on
the $xy$ plane. In the present model, $\phi$ is set to be 0 for the 
two disks. The values of $\theta$ can be different for the two:
$\theta_1$ and $\theta_2$ are $\theta$ for the companion and the primary
galaxies, respectively: we used this notation of $\theta_1$ for the companion,
because we focus on the gas dynamics of the companion.

We mainly  investigate the models with the following
four merger orbital configurations: 
(i) prograde-prograde (`PP') 
model with $\theta_1$ = 35, $\theta_2$ = $45$,
(ii) prograde-prograde (`PR') 
model with $\theta_1$ = 35, $\theta_2$ = $135$,
(iii) retrograde-prograde (`RP') 
model with $\theta_1$ = 35, $\theta_2$ = $45$,
and (iv) retrograde-retrograde (`RR') 
model with $\theta_1$ = 145, $\theta_2$ = $135$.
We investigate two representative cases of orbits, bound orbit
with $e_{\rm o}=0.7$ and hyperbolic one with $e_{\rm o}=1$.
The results of the two cases are not so different in terms of
counter-rotating gas formation in S0s.
The model parameters for each model  are summarized in Table 1.

\section{Analysis}

\subsection{Mapping of Simulated Data}\label{section:mapping}

Here we briefly describe our methods for producing two dimensional (2D)
maps from our three dimensional (3D) SPH simulations. This process is
essential in presenting a clear picture of our simulations, however
the specific details can have some bearing over the interpretation of
our results. Note that, while gas and stars are often treated
separately in our analysis, the mapping process for both components is
identical. 

First, we perform a coordinate transformation of the positions and
velocities of each particle for a given timestep of a given simulation
in order to achieve the desired projection. Typically, this is either
edge on or nearly face on (inclined at 15-20$^{\circ}$) with respect to the disk of the central
galaxy. The reason we do not present perfectly face on projections is
because this will result in an apparent lack of rotation in the
velocity maps for rotating galaxies. This is key to clearly showing
counter-rotating components in our simulated galaxy mergers.

Next, we create a 2D, 100$\times$100 kpc grid with a
pixel size of 0.5 kpc. At each grid point we select all particles
within 4 kpc of the pixel centre. For each selected particle we
calculate a weighting using a Gaussian kernel with a FWHM of 1 kpc
multiplied by the particle mass. From this we produce maps of mass
surface density by performing a weighted sum of the selected particles
and dividing by our pixel size of 0.25 kpc$^{2}$. For velocity maps,
we take the weighted average of the $z$-component of the velocity
vector (projecting into or out of the 2D plane defined by our
grid) for particles selected about each pixel. A caveat to our kinematic
measurements is that we will only trace the dominant kinematic
component at a given spatial location, while information on multiple,
co-located kinematic components is lost. This smoothing is, however,
significantly larger than our softening length, thus we can be sure
that our results will not change were we to run the same simulations
at much higher spatial resolution.

Maps produced using this technique, including surface mass density and
kinematics of different components, are shown in Sections
\ref{section:crcons} and \ref{section:morphs}. Although we do not
suffer from the effects of noise and limits on our sensitivity to low
surface mass density regions, these maps provide a useful qualitative
comparison to observational data products such as those produced from
IFS instruments and even radio interferometry.

\subsection{Integrated Quantities}\label{section:intquant}

\begin{figure}
  \centering
  \includegraphics[width=\columnwidth]{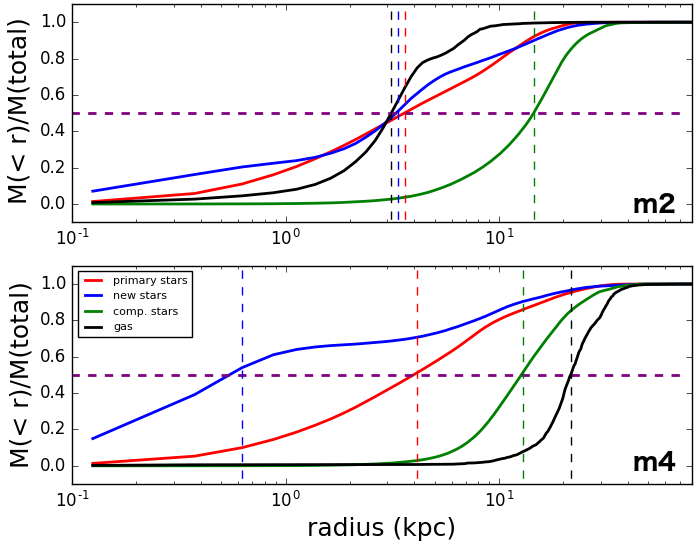}
  \caption{An illustration of our method of measuring the radial
    extent of each component in our simulated merger remnants. Solid
    lines track the fractional mass contained within a given radius
    while vertical dashed lines with corresponding colours mark the half-mass
    radius of a given component. The horizontal dashed line shows the
    value 0.5 for reference.}\label{figure:extex}
\end{figure}

In this section we briefly describe a number of physical quantities measured
in each of our simulations. These include the final half-mass radii and
concentrations of each baryonic component, the percentage of gas
consumed by star-formation, and the percentage change in angular momentum
($L$) for gas and primary galaxy stars. These values are summarised in
the appendix in
Table \ref{table:3}, and we describe below our methods for measuring
these values.

First, half-mass radii and concentration index are measured. We
measure the spherical radius defining spheres in which contain 50 and
90\% of the total stellar mass giving $r_{50}$ and $r_{90}$,
respectively. The concentration index, $c$, is then defined as the
ratio of these values, $r_{90}$/$r_{50}$. This value is similar to the concentration
index employed on observational data, which \citet{nakamura03} and
\citet{deng15} show that $c$ can be a useful 
tool in separating between early- and late-type galaxies. The typical
value separating morphologies is $c$ = 2.85, with early-type galaxies
being more concentrated and late-type galaxies being less
concentrated. We present these values for gas, primary
stars, companion stars, and newly formed stars. An illustration of
this procedure for m2 and m4 is given in Figure \ref{figure:extex}. 

The next integrated quantity is the percentage of the initial gas mass
that is consumed by star formation. This quantity is measured by first
subtracting the final gas mass from the initial gas mass, then
dividing by the initial gas mass. This gives a decimal value that is
then converted to a percentage.

Finally, we calculate the change in angular momentum about the primary
galaxy of the gas
component and the stars initially contained within the primary galaxy
for each model. This is done by calculating the cross product of the
radial and velocity vector of each particle in a given component,
taking the mass weighted center and average velocity as the zero
points of the system. This is then multiplied by the mass of the given
particle, and each angular momentum vector is summed giving $\vec{L_{tot}}$, and the final
value is the magnitude of this vector. This process is
performed at the beginning and end of the simulations and the
percentage change is calculated in a similar manner to the percentage
change in gas mass. 

\section{Results}\label{section:results}

\subsection{Isolated Disk Galaxy Model}

Before examining our minor merger simulations we first briefly explore
the properties of our isolated model m1. This simulation begins as a
featureless disk of gas and stars with a central bulge having a bulge
mass fraction of 0.17. An edge-on view of the initial setup and a
face-on view after 1.4 Gyr of evolution in isolation are shown in the
top two panels of the first column of Figure \ref{figure:kine1}. From
the view of the final gas and stellar distributions, we find that this
galaxy has developed a well defined spiral arm pattern meaning this
simulation can be clearly classified as a late-type galaxy (LTGs). In this
work, we consider the presence or absence of spiral arms to be the key
factor separating LTGs and S0 galaxies. We also note that the final $c$
value (see Table \ref{table:3}) for the stellar component is < 2.85,
consistent with this being a LTG. 

Below the stellar and gas density maps we show the final kinematics
maps of the old stars, new stars, and gas, which are found to share a
very similar disk-like rotation. At large radii, the kinematics of the
old stars exhibit a fairly random distribution, however this is simply
due to the presence of a pressure supported bulge. The compact profile
of this galaxy component has a broad, low density wing of stars
resulting in a small fraction of bulge stars being found beyond the
full extent of the disk, particularly in directions perpendicular to
the disk distribution. As mentioned, these starts are initially
non-rotating, thus in these regions where we primarily find bulge
stars, kinematics are relatively random. 

In our isolated model we have found a strong spiral structure and
co-rotation among all components of our simulated galaxy. Thus, we can
say with confidence that the absence of these features in our
simulations of minor mergers are the result of the galaxy interraction
rather than secular processes. It can be argued that the presence of a
massive gas disk in our isolated model means that this is a somewhat
unfair comparison to our initially gas-free primary galaxies, however
previous works have shown that spiral arms can be formed in
simulations of isolated, gas-free disk galaxies \citep[e.g.][]{bekki02}.

\subsection{Conditions for Producing Counter-Rotating Gas Disks in S0 Galaxies}\label{section:crcons}

%%%%% TABLE2
\begin{table}\label{table:2}
\centering
\begin{minipage}{\columnwidth}
\begin{center}
\caption{Counter-Rotation and/or Rings}
\begin{tabular}{|l|c|c|}
\hline
{ ID \footnote{ The isolated model is m1. 
 }} & 
{ counter? \footnote{ Components with counter-rotation with respect to
      the primary stars\\ in the final merger remnant: c* = companion
      stars, n* = new stars,\\ g = gas, and components with ``!'' are
      only partially counter-rotating\\ (e.g. kinematic twists or flips)
 }} &
{ rings?}\\
\hline
\hline
m1 & n/a & none\\
m2 & c*, n*, g & inner\\
m3 & c*, n*, g & inner, outer\\
m4 & none & inner\\
m5 & none & outer\\
m6 & c*, n*!, g & inner\\
m7 & c*, n*! & none\\
m8 & c*, n*! & inner\\
m9 & c*, n*!, g! & inner, outer\\
m10 & c*, n*, g & inner\\
m11 & c*, n*, g & inner\\
m12 & none & inner, outer\\
m13 & g & inner\\
m14 & c*, n*!, g! & inner, outer\\
m15 & c*, n*, g & inner\\
m16 & c*, n*, g & inner\\
\hline
\end{tabular}
\end{center}
\end{minipage}
\end{table}

In this Section we explore the effects of the relative spins and gas
fractions of galaxies in our merger simulations on the gas versus
stellar kinematics of the merger remnants. Specifically, we are
interested in identifying the key initial conditions that result in a
counter-rotating gas disk. For a quick reference, we provide in the
second column of Table 2 a list of components that are
counter rotating relative to stars originally in the primary galaxy
for each model.

Before presenting our results, we must explicitly state the
definition of counter rotation used in this work. We consider the
primary baryonic component of these galaxies to be the stars initially
belonging to the central galaxy as these stars make up a majority of
the mass budget of our merger remnants. Thus we define a given
component at ``counter rotating'' if it exhibits rotation in the
opposite direction to those stars initially belonging to the primary
galaxy. In IFS observations of merger remnants
such as these, stars initially in the primary
galaxy will represent the primary spectral component. In particular,
for IFS analyses in which only a single stellar component is assumed,
only the most massive kinematic component will be analysed. This is
typically true of shallow observations, while observations with longer
exposure times can reliably extract multiple stellar kinematic
components. In star-forming galaxies, gas kinematics are also often
straightforward to measure from the roughly Gaussian emission lines. Thus gas
versus stellar counter-rotation is often the easiest to identify from
an observational perspective.

\subsubsection{Dependence on Initial Angular Momentum}\label{section:gfamres}

\begin{figure*}
  \centering
  \includegraphics[width=\textwidth]{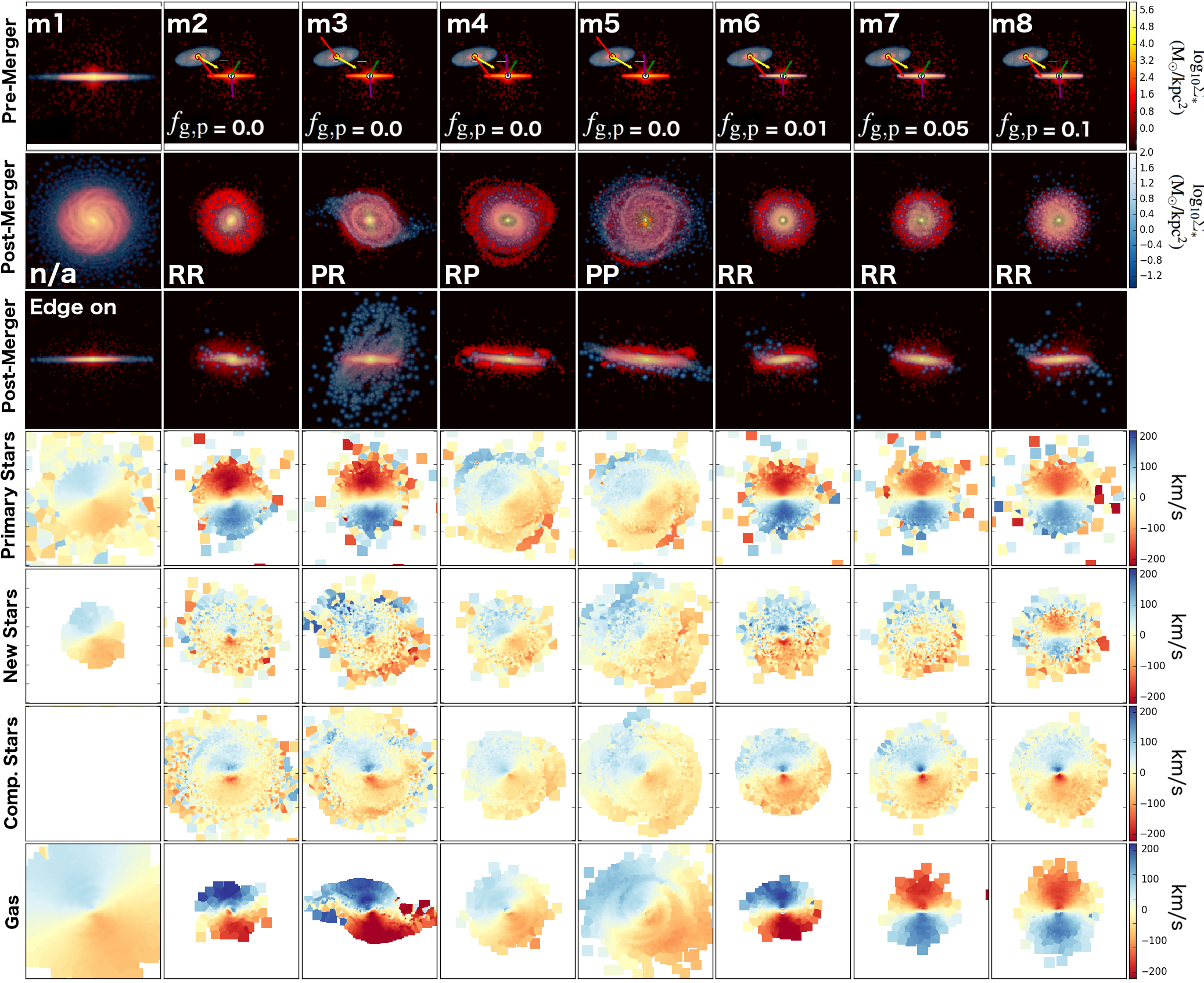}
  \caption{The distribution of gas and stars and stellar and gas
    kinematics for each component for models m1-m8. The top row shows
    the initial edge on view of the primary galaxy, with the incoming
    companion clearly visible. Rows 2-7 show the final conditions of
    the galaxy in the same, nearly face-on view. Rows 2 and 3 show the gas
    and stellar distributions in face and edge-on projections while the remaining rows, from top to
    bottom, show the final kinematics of the primary stars, new stars,
  companion stars, and gas. Models numbers and primary gas fractions
  are shown in the top row and initial relative angular momentum is
  indicated in the second row. Each panel has a size of 100x100 kpc with
the exception of panels for m1, which are 70x70 kpc in order to better
illustrate the spiral arm pattern.}\label{figure:kine1}
\end{figure*}

Here we examine the stellar versus gas kinematics of four simulated
mergers with $f_{g,p}$=0.0, models m2-m5,
beginning with very similar initial conditions. We fix the initial
orbits, $M_{*}$ of both galaxies, $f_{g,c}$,
and rotational speed of both galaxies. The only difference in each
simulation are the direction of the spin of each galaxy with respect
to the spin of the orbit, either retrograde or prograde. Thus, each
simulation may be distinguished by the relative spin of the companion
and the primary galaxies, e.g. retrograde-prograde. For the remainder
of this work, we will refer to the relative spins in a given
simulation in this manner, giving first the spin of the companion
followed by the primary (e.g. ``RP'' denotes a retrograde companion
and a prograde primary). These scenarios are illustrated in the
columns 2-5 of Figure
\ref{figure:kine1} where we show from left to right models
m2 (RR), m3 (PR), m4 (RP), and m5 (PP). 

The first result of these four simulations is that mergers
resulting in counter-rotating gaseous and stellar components require that the primary
galaxy be rotating retrograde to the merger orbit. 
This is consistent
with a number of previous works focusing on both dissipative and
dissipationless major mergers
\citep[e.g.][]{kormendy84,balcells90,hernquist91,bendo00,dimatteo07}. In regards to
counter- versus co-rotation of accreted material, however, the spin of the companion galaxy
has no effect. This is due to the fact that the companion galaxy is
completely destroyed in the merger and its gas and stellar content is
redistributed in the disk of the primary galaxy. In this way the
initial internal kinematics of the companion galaxy are lost and the
orbital direction of the accreted material is inhereted by the orbital
direction of the merger. 

Although the spin of the companion galaxy does not seem to affect the
spin of the gas and new stars of the merger remnant, it does have a
clear effect on the radial extent of the accreted gas. For $f_{g,p}$ = 0
mergers in which the rotation of the companion
galaxy is retrograde, RR (m2) and RP (m4), the gas content of the
merger remnant appears in Figure \ref{figure:kine1} to be less extended
than for our PR (m3) and PP (m5) mergers. This agrees with the
$r_{f50,g}$ values in Table \ref{table:3} with m2 and m4 having values
of 3.1 and 10.4 kpc, compared to 13.1 and 21.6 kpc for models m3 and
m5. Furthermore, the concentration indices of m2 and m4 are $\sim$47\%
larger than m3 and m5, meaning the former two have significantly more
compact gas distributions. The
relative spin of the primary galaxy has a the same effect on the gas
extent, however the effect
is not as large. For example, $r_{f50,g}$ of our RP merger is
3.4$\times$ larger than for our RR, and gas in our PR merger is 4.2$\times$
more extended than in our RR remnant.
Therefor we find that the merger remnant among m2-m5 with the
smallest gas disk is our RR merger while the merger with the largest
gas disk is PP. We discuss further the varying morphologies of each
component in Section \ref{section:morphs}.

\begin{figure*}
  \centering
  \includegraphics[width=\textwidth]{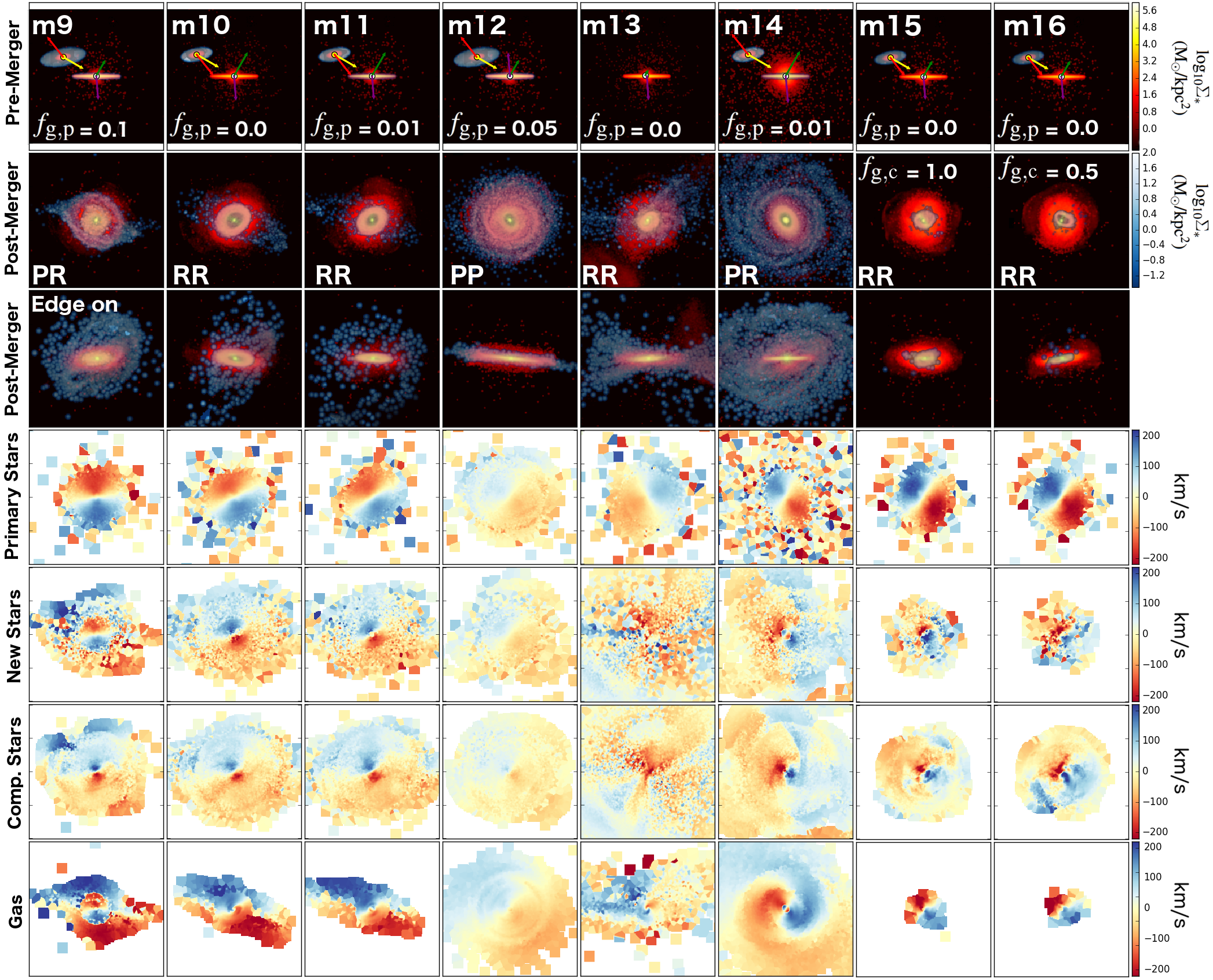}
  \caption{Initial and final stellar and gas mass densities as well as
  final kinematics for each component of our merger remnants, similar
  to Figure \ref{figure:kine1}. In this Figure we explore the effects
  of altering the orbital parameters (e.g. initial galaxy separation
  and eccentricity) and primary bulge mass fraction on the final
  kinematics of our merger remnants.}\label{figure:kine2}
\end{figure*}

\subsubsection{Dependence on Central Gas Mass}

Next we explore how the properties of the merger remnant depend on the
gas content of the primary galaxy. In this Section we take our $f_{g,p}$=0, RR
merger, model m2, from Section \ref{section:gfamres} as our base model and
run an additional three simulations with $f_{g,p}$ increasing each
time. We perform mergers with primary gas fractions of 0.01, 0.05,
and 0.10, which correspond to 0.2 (m6), 1.0 (m7), and 2.0 (m8) times the total gas
mass of the companion galaxy. In each
simulation the gas in the primary galaxy is arranged in an exponential
disk with a scale radius and scale height matched to that of the
stellar disc
component of the primary galaxy. The initial and final gas/stellar
distributions and final kinematics are shown in columns 6-8 of Figure
\ref{figure:kine1}. 

The main effect of adding gas to the primary galaxy is that gas
accreted from the companion during a retrograde merger collides with
gas in the primary. In all four simulations the gas is accreted in
such a way that its bulk motion is opposite to the direction of
rotation of the primary disk. Thus, the rotation of the accreted gas,
which follows the orbital angular momentum of the companion's orbit,
will be decreased relative to the gas free case. This can be seen
Figure \ref{figure:kine1}, comparing to bottom rows of columns 2 and 6
showing the final gas kinematics. The base $f_{g,p}$=0 model, m2,
shows clear gas-stars counter rotation with a complex gas velocity field. The same
is true of m6 with $f_{g,p}$=0.01,
however the irregularities in the gas velocity have been smoothed
out. Our two models in which the primary galaxy begins with the same
or more gas than the companion, m7 and m8, have
co-rotating gas and primary stars in their merger remnants. In
general, accreted gas is swept up by any preexisting gas in the 
primary galaxy. The larger the gas mass initially in the primary, the easier it is for
accreted gas to change its orbital direction. 

We also find that both
the gas and primary stars of merger remnants with non-zero $f_{g,p}$
are more extended than in the $f_{g,p}$=0 case. Increasing $f_{g,p}$
results in an increase of $r_{f50,g}$, with models m6, m7, and m8
having $r_{f50,g}$ = 6.1, 7.6, and 10.6 compared to 3.1 for m2. Adding
even a small amount of gas to the primary galaxy is found to also significantly
reduce $c_{g}$, however, beyond this initial drop,
adding more gas does not seem to further decrease this value. Models
m6-m8 all have similar $c_{g}$ values at $\sim1.6$ compared to
$c_{g}$=2.2 found for m2. We will discuss further the relative
morphologies of different components of each model further in Section
\ref{section:morphs}. 

Considering the kinematics of stars initially belonging to the
companion galaxy, row 5 of Figure \ref{figure:kine1}, we find that
including gas in the primary galaxy has 
little to no effect. There is no appreciable difference between models
m6-m8, which are also quite similar to those of m2, the $f_{g,p}$=0
case. This is because stars merge without dissipation and thus retain
the motion of the companion's orbit about the primary galaxy. The
kinematics of the new stars shown in row 4 of Figure
\ref{figure:kine1}, on the other hand, are somewhat
more complex. In the central regions we find that newly
formed stars are co-rotating with the primary galaxy stars. This is
even true for model m6, although the co-rotating region is extremely
small with a radius of $\sim$3.5 kpc. This co-rotating region
increases in size with increasing $f_{g,p}$. Beyond this inner
co-rotating region, the new stars are found to counter rotate. The
reason for this complex behavior is that stars in the inner region
form from gas that is either initially belonging to the co-rotating
gas disk of the primary or from accreted gas that has reached smaller
radii and is thus more likely to have been swept up by the preexisting
gas. Counter rotating new stars at large radii formed from gas
initially in the companion galaxy before this gas has been swept up by
the gas of the primary. Thus, these new stars
retain the counter-rotating orbits by merging dissaptionlessly.

Finally, we also explore the effect of the initial gas content of the primary
galaxy in a PR merger. This was shown in the previous section to be
the only other configuration resulting in counter-rotating gas for
models with gas free primary galaxies (m3). We show the
final stellar and gas velocity maps for PR mergers with and without
initial gas in the primary galaxy in column 3 of Figure
\ref{figure:kine1} and column 1 of Figure \ref{figure:kine2},
respectively. The overall spatial distributions of gas and stars, as well
as the stellar kinematics, in the two models are quite similar, however
there is a clear difference in the gas kinematics. In m9 we find
that in the central region gas and stars are co-rotating while gas
accreted at larger radii is counter-rotating as in m6. This is in
contrast to nonzero $f_{g,p}$ models with RR orbits, m6-m8 discussed
above, which exhibit smooth velocity maps with no kinematic flips.
As we noted in the previous section, mergers
with a prograde satellite rotation result in a larger final gas
disk. In the case of m6-m8 all the accreted gas falls to small radii and
is swept up by the initial gas content of the primary galaxy, while in
m9 gas rapidly stripped from the companion collects at radii beyond the
central gas disk and is thus able to retain the counter-rotating
kinematics inherited from the orbit of the merger. 

\subsubsection{The Effect of Orbit, Bulge Mass, and $f_{g,c}$}

Next we briefly investigate the effect of the orbital distance and
eccentricity, primary bulge mass, and $f_{g,c}$ on the final
distribution and kinematics of gas and stars in our merger
simulations. We show the initial setup, the near face-on view of the
final merger remnant, and the final near face-on stellar and gas
kinematics of these models in the columns 2-8 of  Figure
\ref{figure:kine2}.

Considering merger orbit, we present two RR simulations that both have orbits
with initial and pericenter separations between the two galaxies twice
that of our previous models. These models are m10 and m11 in Table
1, and the gas fractions of the primary galaxies are 0 and
0.01 respectively. We also produce a PP merger with an initial
separation 2$\times$ larger than in previous sections and with
$f_{g,p}$ = 0.05, m12. Finally, we present model m13, which has an
orbital eccentricity of 1.0, a parabolic orbit, requiring a
significantly larger initial separation to achieve the same pericentre
distance as our other models. 

For models m10 and m11, the resulting merger remnants are quite
similar to m2, exhibiting counter-rotating
gas in the central regions. In contrast to m2, however, both models
have a more prominent ring structure and long tidal streams of gas
with a slightly different orbital
plane. These tidal streams appear to have inherited their orbits from
the initial orbit of the incoming companion galaxy, and in the final
kinematic maps their rotation is offset by $\sim$10-30$^{\circ}$. This is
consistent with the inclination of the orbit we input as initial
conditions. As with previous simulations, the kinematics of new stars
tend to follow those of the gas, and the kinematics of stars initially
in the companion are largely unaffected appearing quite similar to
models m2-m8. Model m12, our PP merger, has resulting gas/stellar
distributions and kinematics that are extremely similar to the
$f_{g,p}$=0, PP merger, m5. Both m5 and m12 are among the models with
the largest $r_{f50,g}$ at 21.6 and 16.9 kpc, respectively. The radial
distribution of the gas is similar as well, with $c_{g}$ values found
to be 1.50 and 1.68. Furthermore, all components in these models are
found to be co-rotating with fairly regular velocity maps. This
suggests that PP mergers may be among the most difficult to identify
based on the kinematics of their remnant galaxies. 

\begin{figure*}
  \centering
  \includegraphics[width=\textwidth]{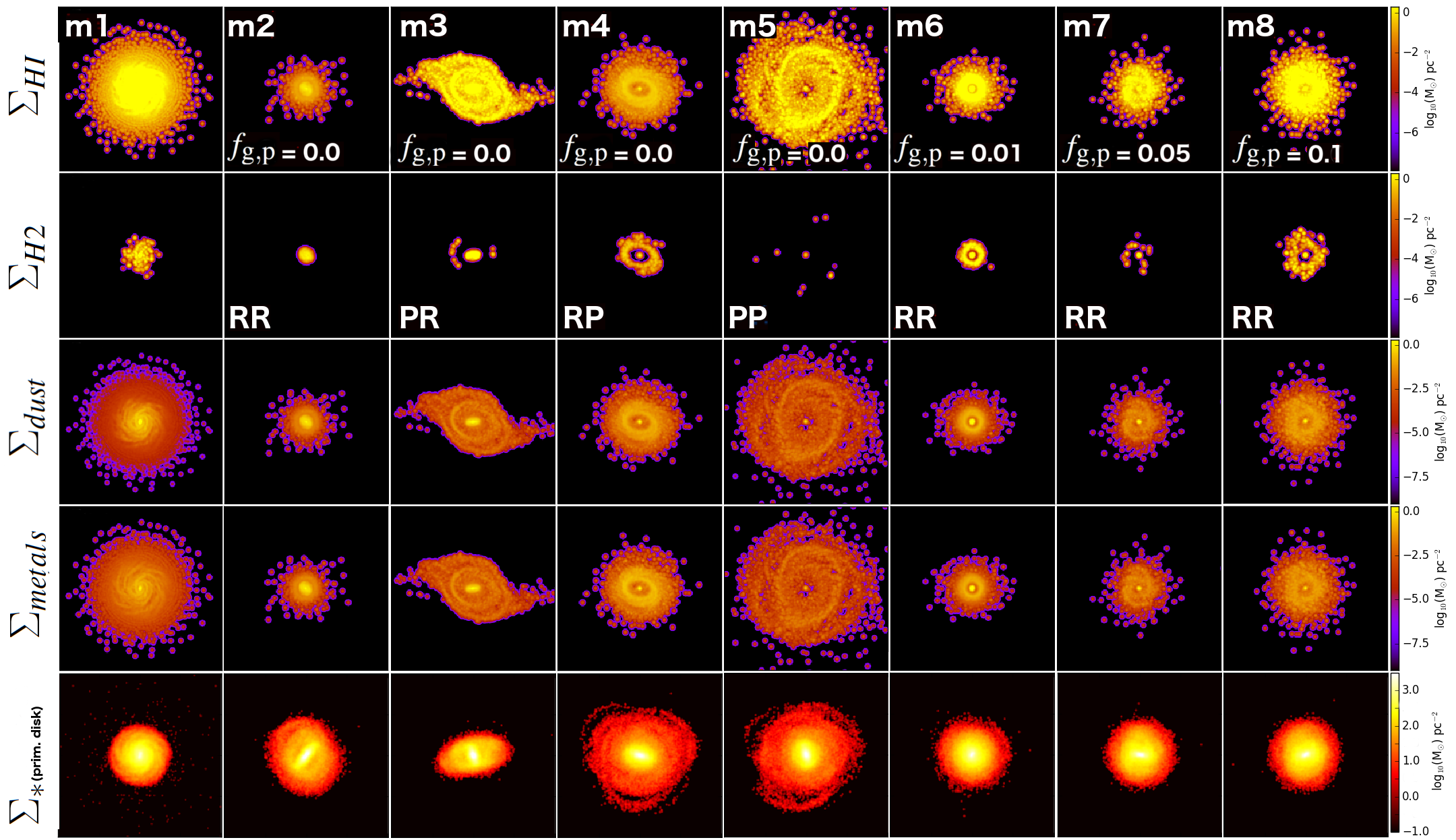}
  \caption{Here we show the final, face-on mass surface densities,
    $\Sigma$s, of, from top to bottom, HI,
H$_{2}$, mass of dust, mass of metals, and stars tagged as initially
belonging to the disk component of the primary galaxy. The reason we
show the $\Sigma_{*}$ map for only those
    stars beginning in the disk of the primary is that it is in this
    component that we find a bar may be induced by the galaxy
    interaction. The first column shows our fiducial/isolated model,
    columns 2-5 show $f_{g,p}$=0 mergers, and
    columns 6-8 depict mergers with increasing primary gas
    fraction. Each panel has a size of 100x100 kpc and each row shares
  a scaling indicated by the colourbars on the right.}\label{figure:gns1}
\end{figure*}

In contrast to all models discussed to this point, model m13 (shown in
column 5 of Figure \ref{figure:kine2}, which
has a parabolic merger orbit, exhibits chaotic kinematics in
the accreted stars and gas. The final distribution of gas has a
centrally concentrated region surrounded by a handful of long streams
that roughly mirror the distributions of accreted stars. These streams
exhibit distinct kinematics that are apparent in the kinematics maps
of the new stars and companion stars. Relative to these two kinematic
maps, the kinematics of the accreted gas is fairly regular, particularly in the
central regions where it counter rotates relative to the primary
stars (which have relatively regular rotation). From an observational
perspective, this galaxy would likely be identified as having gas
versus stellar counter rotation, thus this suggests a parabolic orbit
will not affect the emergence of counter rotation. We note, however
that the random motions of the
accreted stellar components could complicate their
identification in a spectral decomposition of the stellar continuum. 

We also test the effect of significantly increasing the bulge mass of
the primary galaxy in m14, thus testing the effect of a PR wet, minor merger with
an elliptical galaxy. This is also depicted in Figure
\ref{figure:kine2} in the 6$^{\textrm{th}}$ column. The strong
gravitational potential of the massive bulge results in much more
efficient stripping of gas and stars from the incoming companion galaxy. Thus,
the merger remnant of m17 is among the most extended gas disks in our
our simulations with $r_{f50,g}$=21.1 kpc. While the final kinematics
of this model are quite
complex, in the central regions, where IFS observations are
typically focused, the gas is predominantly counter-rotating. In the
inner-most 2 kpc we find a region of co-rotating gas, however this gas
is the remnant of the initial gas disk included in the primary galaxy
in this model. As in previous models, m6-m8, tidal
interactions cause this initial gas disk to collapse inward and be
rapidly depleted through star formation. Beyond the inner regions of
the galaxy, the gas exhibits a kinematic twist that manifests as a
spiral pattern in the velocity map. This is also reflected in the
outer stellar kinematics, which will again be dominated by newly
formed stars similar to models m10 and m11. Similar to previous
models, these stellar motions will
be difficult to detect, but the kinematic twist seen in the gas
content could possibly be detected using radio interferometry for very
nearby galaxies. 

Finally, we test the effects of lowering $f_{g,c}$ on the resulting
properties of our merger simulations. In simulations discussed thus
far we have employed $f_{g,c}$ = 5.0. Although low mass galaxies
with gas fractions this large have been observed, they represent
the most gas rich dwarf systems at low redshift \citep{huang12}. As
such, wet, minor mergers with companion galaxies of
lower $f_{g,c}$ are more common. We have performed two additional
RR, minor merger simulations, m15 and m16, with orbits matched to m2 and $f_{g,c}$ =
1.0 and 0.5, respectively. These are shown in columns 7 and 8 of
Figure \ref{figure:kine2}. Overall the resulting merger remnants are
quite similar to that of m2, featuring counter-rotating gas and
companion/new stars. The radial extent of the gas for all three models is
comparable with m2, m15, and m16 having $r_{f50,g}$ = 3.1, 5.4, and
6.4 respectively. Models m15 and m16, however, have lower $c_{g}$ than
m2 ($\sim$1.4 vs 2.2). It is not clear, though, if this is due to the
change in $f_{g,c}$, or a result of a truncation of the gas disks of
m15 and m16 due to the combination of our mass resolution and the low
space density of gas at large radii in these models. Using a larger
value of $f_{g,c}$ allows us to better map the final distribution of
gas, particularly at large radii, and should not have a major effect
on our main conclusions regarding the emergence of counter-rotation. 

\subsection{Merger Remnant Morphologies}\label{section:morphs}

In all of our simulations, the merging process results in a S0 merger
remnant, i.e. a disk galaxy devoid of spiral arms. This is in contrast
to many previous works on simulations of counter-rotating gas and
stellar components in merger
remnants focusing on major mergers as these simulations typically
result in elliptical morphologies
\citep[e.g.][]{dimatteo07,jesseit07}. In this sense, the mergers
presented here provide an important step forward in understanding the
formation mechanism for the 20-40\% of S0 galaxies found to host
counter-rotating components \citep{pizzella04,davis11}. 

We note that in some models
the images of the final morphologies shown in the second rows of
Figures \ref{figure:kine1} and \ref{figure:kine2} appear as late
types, models m12 and m14 in particular. This appearance is driven by
the gas distributions at large radii, rather than stars. To show this
we create simulated SDSS r-band images of model m17 placed at a
redshift of 0.05. First an artificial datacube was created using the
simple stellar population spectral models from the P\'{E}GASE-HR library
\citep{leborg04}, created from observations of Milky Way stars. In
each spaxel of our simulated cubes we determine the age of each star
particle based on their time of birth and the time elapsed in the
simulation. Star particles present at the beginning of the simulation
are initiated with an age of 1 Gyr. We then create a mass weighted
spectrum in each spaxel where each star particle contributes a
component taken as the P\'{E}GASE-HR model with an age closest to that
of the particle. We next blur the image using a Gaussian kernel with
a full-width half max of 3.61 pixels, corresponding to the average
seeing of the SDSS survey. Finally, we apply the SDSS r-band filter to this cube
and add Gaussian noise at a level that produces a simulated image
typical of SDSS observations at $z=0.05$.

\begin{figure}
  \centering
  \includegraphics[width=\columnwidth]{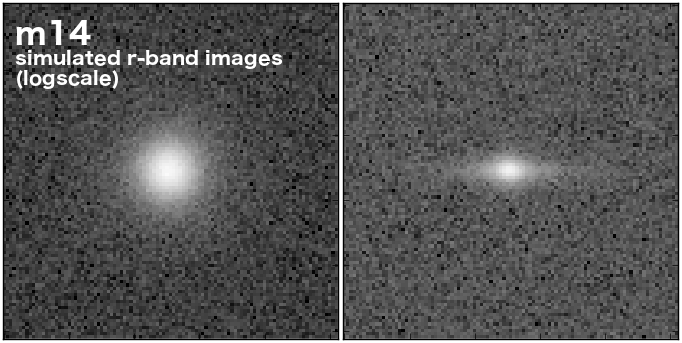}
  \caption{Face-on (left) and edge-on (right) simulated r-band images
    of model m17 produced from an artificial datacube as described in
    Section \ref{section:morphs}. The galaxy is assumed to be at a
    redshift of 0.05, and a Gaussian blurring matching the average
    SDSS seeing has been applied. We also add Gaussian noise at a
    level chosen to roughly match the appearance of SDSS galaxies at
    $z$=0.05.}\label{figure:rband}
\end{figure}

Face-on and edge-on simulated r-band images for model m17 are shown in
Figure \ref{figure:rband}. This shows that the spiral arms apparent in
Figure \ref{figure:kine2} are not present in the stellar distribution,
which has a lenticular appearance. The gas component of these galaxies
at large radii is almost entirely HI, thus star-forming regions will
not be present. This means that even for photometry at shorter
wavelengths, e.g. the u-band, will not provide imaging of the gas.

\begin{figure*}
  \centering
  \includegraphics[width=\textwidth]{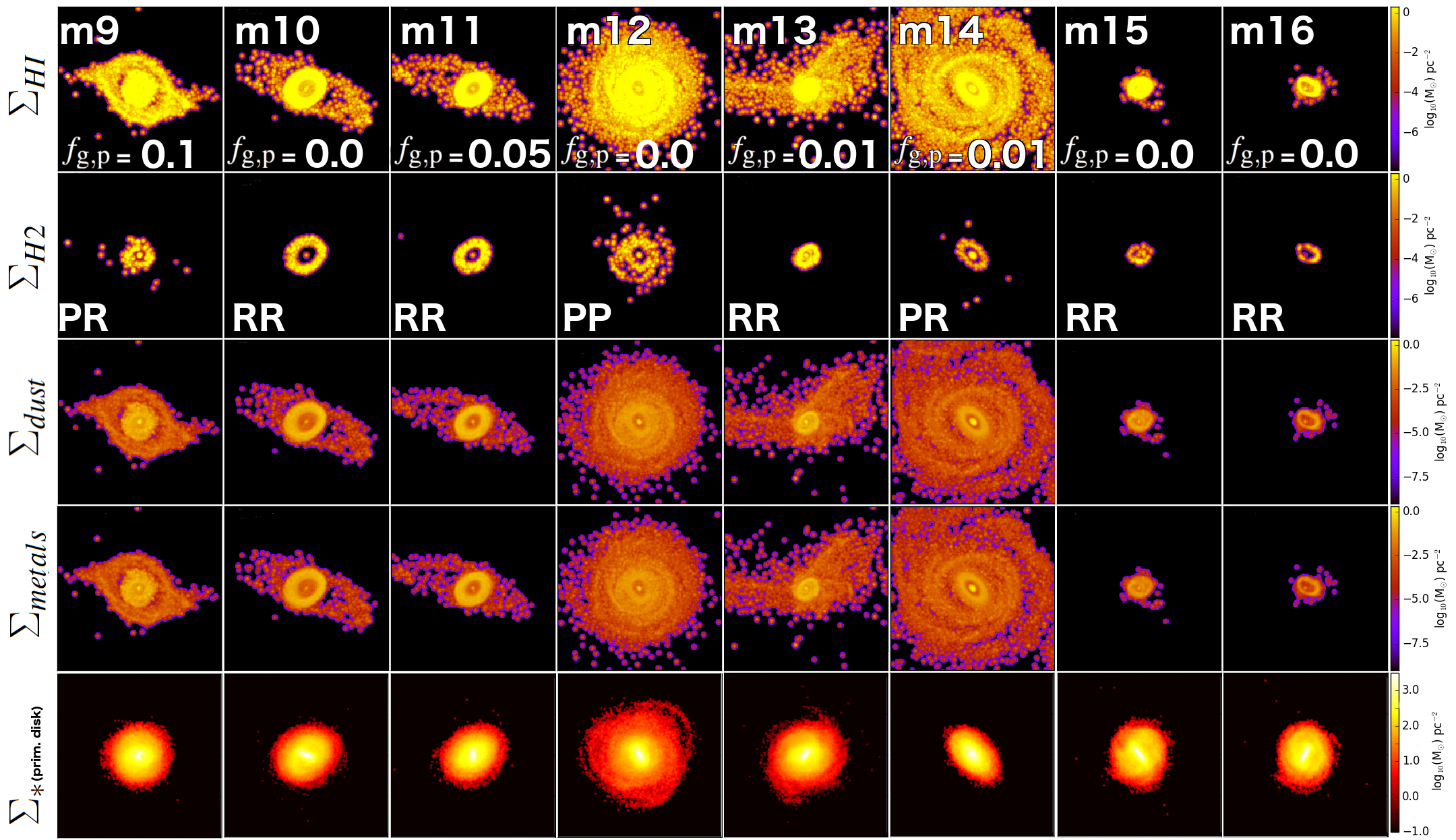}
  \caption{Rows, image sizes, and scale bars here are the same as in
    Figure \ref{figure:gns1}, however now we show results for models
    9-16. In these models we modify the merger orbit, bulge
    fraction, and $f_{g,c}$ as discussed in the previous Section.}\label{figure:gns2}
\end{figure*}

In a majority of our merger simulations, the
concentration index of the primary stars in our remnant, $c_{p*}$, is
larger than 2.85. The value of $c$ = 2.85 is found by \citet{nakamura03}
and \citet{deng15} to reliably separate visually-classified LTGs ($c$
< 2.85) and ETGs ($c$ > 2.85). The concentration of new stars,
$c_{n*}$, also tends to be quite high in most cases as a majority of
stars are formed in the inner most regions of the primary
galaxy. These new stars are extremely bright relative to older stellar
populations and will thus contribute to increasing the overall $c$ of
the full stellar distribution. We note, however that in no model is
$c_{g}$ or $c_{c*}$ larger than 2.85, meaning that these accreted
components are preferentially arranged in a disk structure. Thus, from
a quantitative point of view, merger remnants with $c_{p*}$ < 2.85 may
remain as LTGs rather than transforming into S0's. From our simple
prescription in which an S0 is identified as such based on the lack of
spiral arms, however, all galaxy mergers result in S0 remnants as
shown by the second rows of Figure \ref{figure:kine1} and
\ref{figure:kine2}. 

Next we examine in detail the spatial distribution of the various
components of the gas in our simulated merger remnants. We show in the rows of Figures \ref{figure:gns1} and
\ref{figure:gns2} the face
on views of the surface density, $\Sigma$, of, from top to bottom, HI,
H$_{2}$, mass of dust, mass of metals, and stars tagged as initially
belonging to the disk component of the primary galaxy. We observe both
inner and outer rings in our 
simulations, and our large range of initial conditions allow us to
discuss the likely formation mechanism for these structures. The
occurance of rings in each of our models is summarised in Table
2 for reference. Inner rings have typical sizes comparable to
inner galaxy structures such as bars while outer rings are two or more
times larger \citep{athanassoula09}. In this work we employ a fixed
radial cut of 10 kpc to separate inner and outer rings, i.e. rings
with radii < 10 kpc are inner rings and those with radii > 10 kpc are
outer rings. In our
isolated fiducial model, no bar
emerges through secular processes. Thus, we
can say with confidence that bars evident in our merger simulations
are indeed induced by the merging process rather than internal/secular
evolution.

First we examine the formation of inner rings, which occur in eleven
out of sixteen simulations
presented in Figures \ref{figure:kine1} and \ref{figure:kine2}; m2,
m3, m4, m6, m8, m10, m11, m12, m13, m14, m15, and m16. This includes
simulations in which the primary galaxy
initially contains no gas as well simulations in which the primary
galaxy contains gas. Comparing the
morphologies of the gas and stars in our merger remnants, it appears
that the occurrence of inner rings in mergers with $f_{g,p}$
is related to the generation of a bar
in the disk component of the primary galaxy, as seen in the bottom row
of Figures \ref{figure:gns1} and \ref{figure:gns2}. This bar formation
is due to tidal torques
on the disk stars during the merging process and the inner gas rings
in these simulations form at radii consistent with the lengths of the
stellar bars. Inner rings appear to always contain HI

We also find that including large amounts of
gas in the primary galaxy suppresses merger induced bar formation, and
thus the emergence of an inner gas ring at the outer edge of the
bar. The apparent ring structures in models m6, m8, m12, and m14 which initially
contain gas in the primary galaxy, are significantly more compact than
those seen in m2, m3, and m4. This may suggest that the ring-like
appearance in these simulations emerges due to the rapid depletion of
gas in the inner-most regions rather than bar driven resonances. 

Outer gas rings are only prominent in four out of sixteen of these
simulations, always occurring in those simulations in which the rotation of the
companion galaxy is prograde relative to its orbit about the
primary (RP and PP mergers). As we noted in Section \ref{section:gfamres}, the gas
initially in the companions in these mergers is more efficiently
stripped resulting in a much more extended gas distribution when
compared with mergers with retrograde companion rotation. One
difference between our RP and PP mergers is that we find a merger
induced bar in the RP mergers, which is responsible for the generation
of an inner ring in addition the the outer ring (m3 and m9). In the PP case, we
find neither a bar nor an inner ring. We do find gas well beyond
the outer ring, however, whereas in our RP mergers the gas is rapidly truncated
beyond the outer ring. This may be the result of angular momentum and
material being transported inwards due to the influence of the central
bar. The gas in m3 loses 47.9\% of
its angular momentum while for m5 this value is only 12.3\% in
agreement with this assessment (see Table \ref{table:3}). A similar comparison between m9 and
m14, however, is complicated by a 10$\times$ difference in
$f_{g,p}$.

The size of the outer ring in all cases is 
roughly the same, and this may be related to the orbit of the
companion galaxy, which does not change between these
simulations. Further simulations with a larger variation in orbital parameters will
be required to test this. Finally, we note that all of the
outer rings observed in our simulations are almost entirely devoid of
H$_{2}$. Thus, these regions will have a large hydrogen mass with
relatively low levels of star-formation, and should therefore
represent outliers from the kennicutt-schmidt relation
\citep{schmidt59,kenni98}.

Although we do not observe spiral arms in any of our merger remnants,
we do observe arm-like tidal streams in three of our
simulations. Specifically these are models m4, m5, and m12, which are
also those three simulations with a prograde primary
galaxy. Furthermore, these are also the only three galaxies that
exhibit no counter-rotating components in their merger
remnants. Although observed low surface brightness features such as
these often requires long exposure times, these may be useful in
identifying merger remnants with prograde primary galaxies that will
not display clear kinematic signatures.

Before leaving the subject of merger remnant morphologies we also
highlight another peculiarity that emerges in models m2 and m3. These
models are both performed with retrograde primary galaxies and
$f_{g,p}$ = 0, resulting in the two strongest stellar bars among all of our
models (see the bottom row of Figure \ref{figure:gns1}). We find that
these models also exhibit bars in their distributions of gas, dust,
and metals, and this bar is significantly shorter and perpendicular to
the stellar bar. These so-called ``nested'' bars were originally
proposed as a mechanism for fueling active galactic nuclei by
\citet{shlosman89} and have subsequently been observed in around 1/3
of barred galaxies \citep[e.g.][among
others]{wozniak95,erwin02,erwin04}. Works on simulations of nested
bars are on-going with some groups manually creating these structures
to explore their influence on the host galaxy evolution
\citep[e.g.]{namekata09} or, notably, \citet{wozniak15} who
demonstrate the emergence of nested bar formation in an isolated disk
galaxy. Our simulation is unique as this serendipitous discovery of a
nested bar that appears to be induced by a galaxy merger. The topic of
merger induced nested bars will be the topic of future research, and
we simply make note of it here.

\subsection{Other Galaxy Properties}

Here we present the evolution of star-formation and dust scaling
relations of our merger simulations. While not directly related to the
emergence of counter-rotation in S0 galaxies, these provide
signatures that can be reliably compared with current observations.

\subsubsection{Star-Formation}

\begin{figure}
  \centering
  \includegraphics[width=\columnwidth]{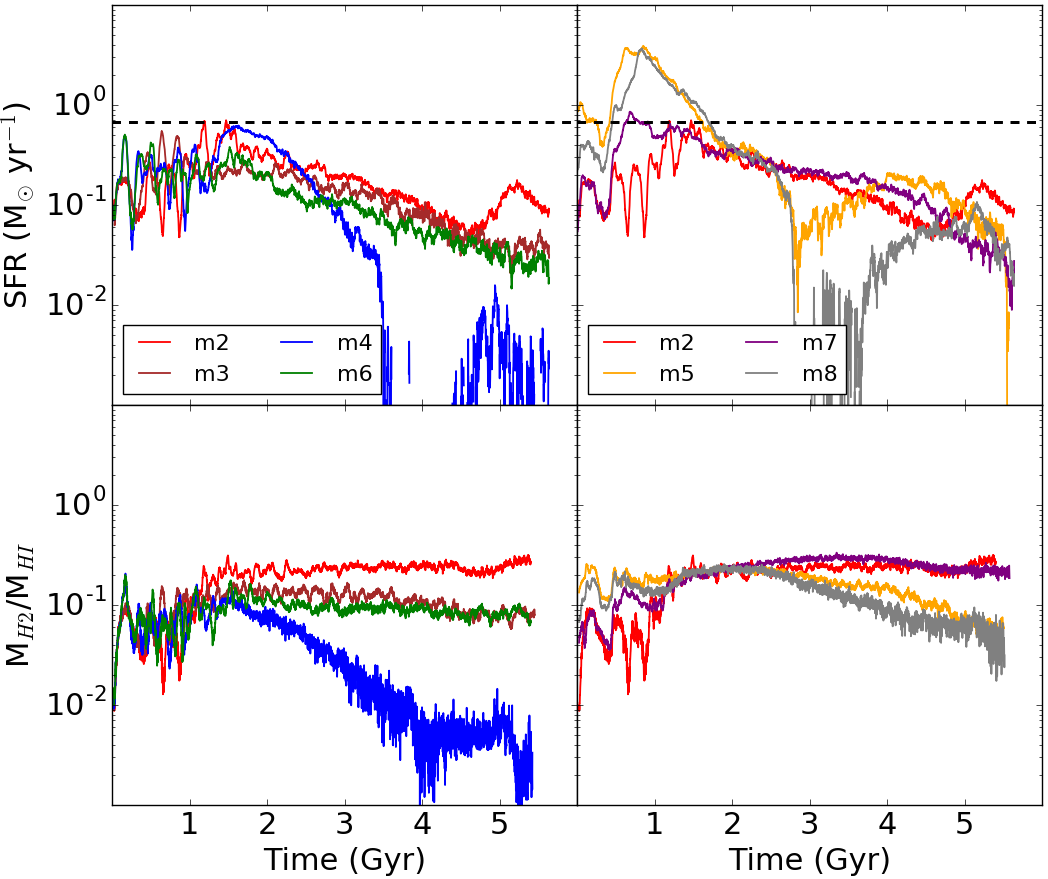}
  \caption{The time evolution of SFR and H$_{2}$ gas mass fraction for
  models m2-m8. Models with gas free primary galaxies are shown in the
left column while those initially containing gas are shown in the
right. Note that our gas-free primary, RR merger, m2, is shown in both
columns for comparison. In the top row we also indicate with a
horizontal dashed line the expected SFR for a main-sequence,
star-forming galaxy at the mass of our primary galaxies assuming the
relationship found by \citet{whitaker12}.}\label{figure:gfevo1}
\end{figure}

\begin{figure}
  \centering
  \includegraphics[width=\columnwidth]{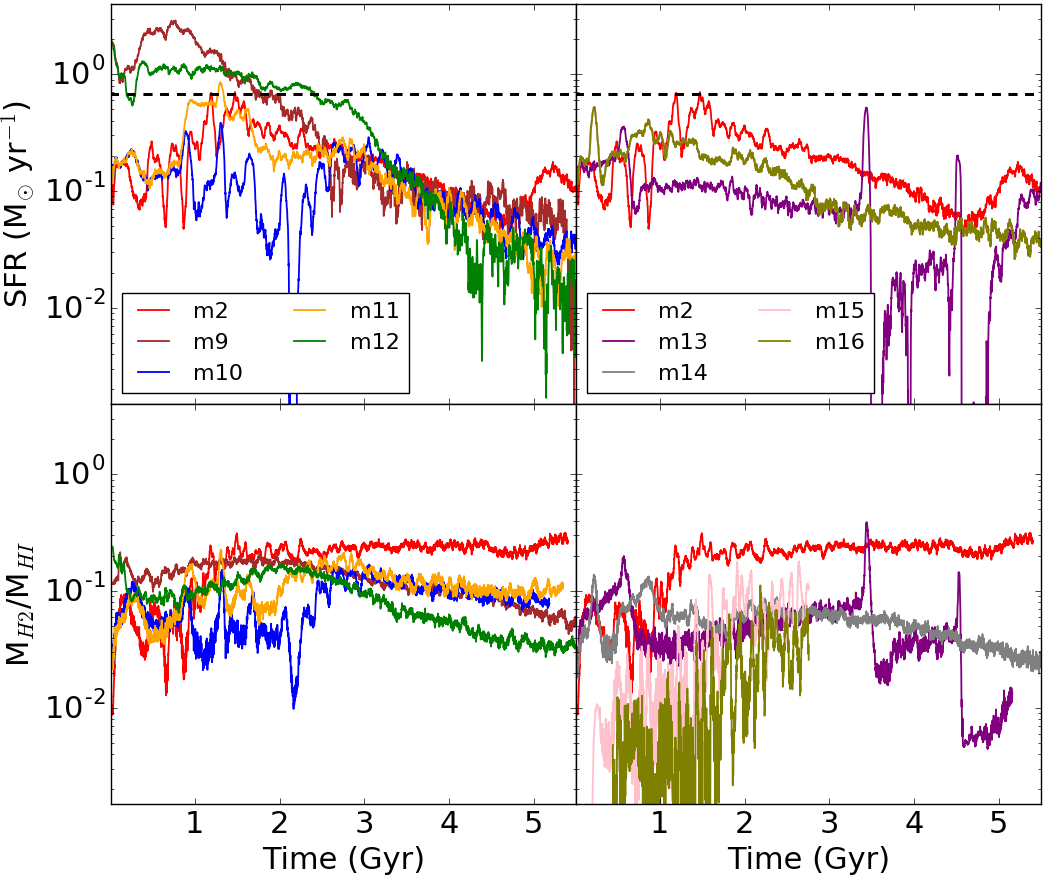}
  \caption{The time evolution of SFR and H$_{2}$ gas mass fraction for
  models m9-m16. Models with an increased initial orbital distance are
  shown in the left column while the right column shows our parabolic
  orbit model (m13), our large bulge model (m14), and our low
  $f_{g,c}$ models (m15 and m16). We also show in both columns our
  $f_{g,p}$ = 0, RR merger (m2) for comparison. As in Figure
  \ref{figure:gfevo1} we include a
horizontal dashed line showing the expected SFR for a main-sequence,
star-forming galaxy at the mass of our primary galaxies.}\label{figure:gfevo2}
\end{figure}

The time evolution of SFR and H$_{2}$ fraction ($M_{H2}/M_{HI}$) is
shown for  m2-m8 in Figure \ref{figure:gfevo1}. The top row shows the evolution
of SFR and the bottom row shows the evolution of H$_{2}$ fraction. The
horizontal dashed line in the top row indicates the SFR of the
$z=0$ star-forming main sequence at the mass of our merger simulations
\citep[following the prescription of][]{whitaker12}. 

First we focus on our models with $f_{g,p}$ = 0 presented, shown in the left column of
Figure \ref{figure:gfevo1}. We find that all four mergers peak in
their SFR around 2 Gyr, which corresponds to the final coalescence of
our mergers, and this peak is roughly consistent with the SFR expected
for main-sequence galaxies of this mass. Beyond 2 Gyr, the SFR
gradually tapers off in most cases, however m5, our PP merger,
experiences a rapid truncation in SFR at around 3.5 Gyr. The H$_{2}$
fractions of our models gradually increase to 2 Gyr then, in all cases
other than m5, the H$_{2}$ fraction levels off for the remainder of
these simulations. This suggests a coevolution of HI and H$_{2}$ in these
mergers where the H$_{2}$ consumed by star formation is continuously
replenished by the HI reservoir. 

Model m5, on the other hand, was found to exhibit a much
larger spatial distribution of gas due to more efficient stripping
during the merger as a consequence of the PP configuration. Indeed,
Table \ref{table:2} shows that $r_{f50,g}$ for m5 is 21.1 kpc,
significantly larger than the values of 2.1, 13.1, and 10.4 found for
m2, m3, and m4. Gas at
large radii remains stable due to a rapid rotation,
preventing the HI from collapsing to form H$_{2}$. The rapid truncation of SFR seen at 3.5 Gyr is the
result of the depletion of the inner H$_{2}$ reservoir, which is no longer
replinished from the large HI disk. This is also reflected in the
significantly lower H$_{2}$ fraction seen in this galaxy beyond 2
Gyr. 

In the right column of Figure \ref{figure:gfevo1} we show the time
evolution of SFR and H$_{2}$ fraction for our series of RR mergers
with increasing primary gas fraction, m6-m8. Here, model m2 is plotted again as a
reference. In the case where gas is included in the primary galaxy we
find an earlier peak in the SFR that can be slightly above the average
SFR of the main sequence. This occurs because the initial gas disk in the
primary galaxy experiences a tidal torque due to the incoming
companion that causes it to collapse inward. This compaction of the
gas causes an increase in the H$_{2}$ fraction, and as a result the
initial gas in the primary galaxy is rapidly converted into stars. In
the case of m6, there is significantly less initial gas in the primary
and thus the
SFR peak is lower than models m7 and m8. In models m7 and m8, however
we see a rapid truncation of the SFR reminiscent of that seen in our
PP merger, m5. Similar to m5, m7 and m8 result in co-rotating remnants
that have a significantly larger final gas disk. Thus, the truncation
of SFR can again be attributed to stable HI gas at large radii. This
is again reflected in the lower H$_{2}$ fraction when compared with
models m2 and m6. Unlike m5 however, m7 and m8 eventually increase
their SFR again to roughly match the level seen in m2 and m6 at the
end of our simulations. The presence of remaining gas from the primary
galaxy provides a mechanism for angular momentum transfer, allowing
accreted gas to continue to move inwards and form stars.

SFR and H$_{2}$ fraction evolution for models m9-m16 are also shown in
Figure \ref{figure:gfevo2}. The same overall trends observed in models
m2-m8 are also at play here. In general, models in which gas is found
to be trapped at large radii, whether in a disk (e.g. m12 and m14) or
in tidal streams (e.g. m10 and m11) exhibit a suppression of both SFR
and H$_{2}$ fraction compared to m2 at late times. Again this is due to the fact that
gas at large radii is stable and prevented from converting from HI to
H$_{2}$. Also similar to m2-m8, models in Figure \ref{figure:gfevo2}
that begin with a large value of $f_{g,p}$ (e.g. m9 and m12) are found
to have an early peak in SFR related to the rapid depletion of gas
from the primary galaxy. We include Figure \ref{figure:gfevo2} for
completion, however, as star formation is not the focus of this
work we leave in depth analysis for future work.

\subsubsection{Dust Mass Relations}\label{section:dustmass}

\begin{figure}
  \includegraphics[width=\columnwidth]{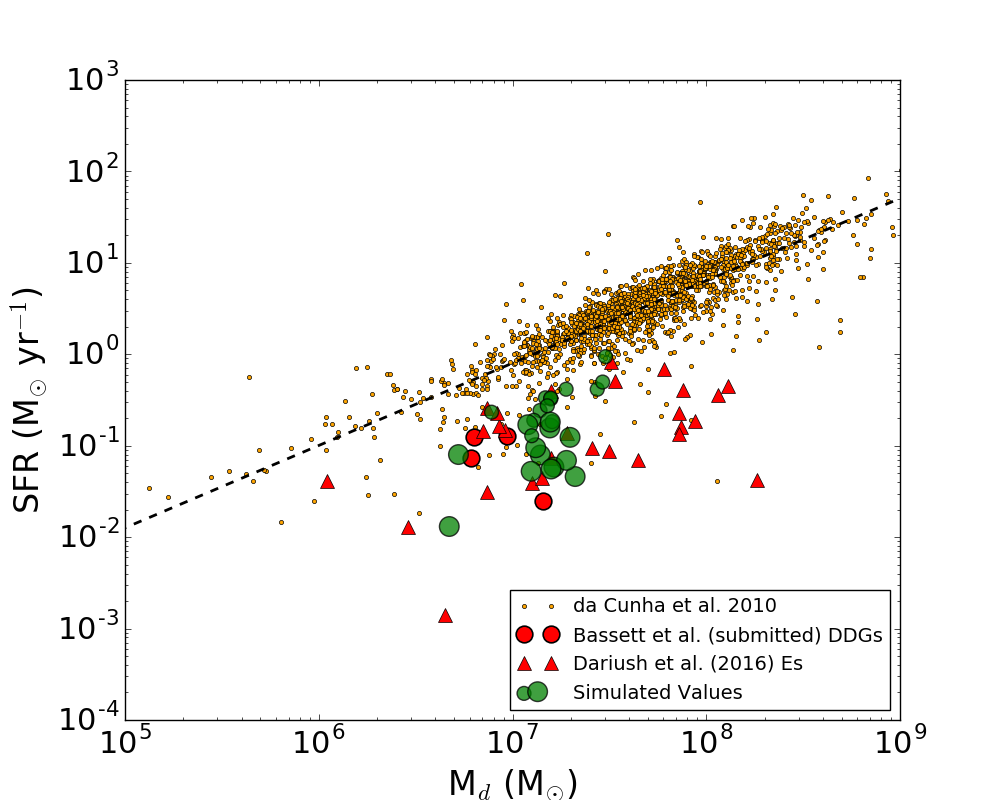}
  \caption{The relationship between $M_{d}$ and SFR for our simulated
    merger remnants. Normal star-forming SDSS galaxies from
    \citet{dacunha10a} follow a tight relationship 
    between $M_{d}$ and SFR and the linear fit to these datapoints is
    shown by the dashed black line. We also show as red triangles and
    circles ETGs containing
    dust from \citet{dariush16} and dusty, dispersion dominated
    galaxies (DDGs) from \citet{bassett17}. Smaller green
    circles show $M_{d}$ vs $M_{*}$ at the halfway point of our
    simulations while larger green circles show the final values.}\label{figure:mdsfr}
\end{figure}

Finally, we explore the dust mass, $M_{d}$, scaling relations of our
minor-merger remnants. Relations between $M_{d}$ and quantities such
as SFR and $M_{*}$ have been explored by a number of authors. In
particular, works such as \citet{dariush16} and \citet{bassett17} use samples of massive visually classified ETGs known to contain dust
based on detection at infrared wavelengths. Galaxies from \citet{bassett17} are further classified as being largely dispersion
supported (``dispersion dominated galaxies'', DDGs) based on IFS
observations from the SAMI Galaxy Survey \citep{bryant15}. The
presence of dust in these
systems is peculiar as these galaxies are likely to host a hot, X-ray
emitting halo that is inhospitable to long lived dust grains
\citep[e.g.]{draine79}. Galaxies from \citet{dariush16} and \citet{bassett17} are found to be outliers in $M_{d}$ vs SFR and $M_{d}$
vs $M_{*}$, which is taken as evidence that their dust content was
recently accreted through mergers. As we have described, our models
carefully track the evolution of $M_{d}$ allowing us to test the
connection between the $M_{d}$-SFR and $M_{d}$-$M_{*}$ relation and
minor-merger activity.

We show in Figure \ref{figure:mdsfr} the relationship between $M_{d}$
and SFR for our minor-merger simulations in comparison with
observations. The relationship for ``normal'' star-forming galaxies
from the Sloan Digital Sky Survey (SDSS) taken from \citet{dacunha10a}
are shown as small orange circles, and a linear fit to this data is
indicated by the black dashed line. Dusty elliptical galaxies from
\citet{dariush16} and DDGs from \citet{bassett17} are shown as red
triangles and red circles, respectively. We show the values for our
merger simulations using green circles at two different time
steps. Smaller circles show the position at the halfway point of our
simulations, slightly after final coalescence, and the larger circles
show the final positions. 

Both our simulated galaxies and dusty ETGs of \citet{dariush16} and
\citet{bassett17} exhibit significantly lower SFRs at a fixed
$M_{d}$ when compared to the bulk of star-forming galaxies from
SDSS. We also find that the SFR of our simulations is typically higher
at the halfway point, which should be expected given the declining
star-formation histories of our models shown in Figures
\ref{figure:gfevo1} and \ref{figure:gfevo2}. It can be seen that a
non-negligible number of star-forming SDSS galaxies also occupy the
low SFR region below the dashed line however, which suggests that
mergers are not the only mechanism that can cause galaxies to fall
below the bulk relation. The fact that all of our simulated data
points fall below this relation is intriguing though, and observations
of $M_{d}$ vs SFR can provide an indication of merging activity,
however further observations will be required to confirm this. Indeed,
IFS observations of the four galaxies from \citet{bassett17} show these
galaxies to have kinematic signatures of mergers such as offsets
between the kinematic position angles of gas and stars
\citep[e.g.][Bryant et al. in prep]{davis13}. 

\begin{figure}
  \includegraphics[width=\columnwidth]{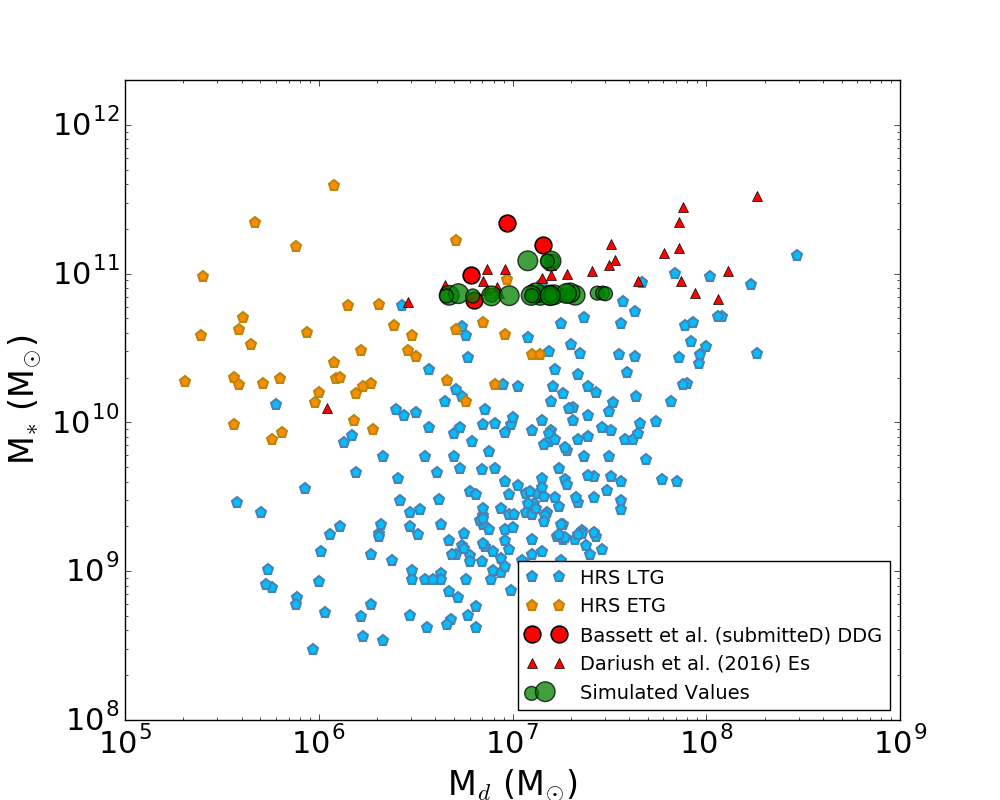}
  \caption{$M_{d}$ versus $M_{*}$ for simulated galaxy mergers, with
    green and red symbols matched to those in Figure
    \ref{figure:mdsfr}. Here we also
    compare with very nearby galaxies from
    the Herschel Reference Survey \citep[HRS;][]{boselli10} with LTGs
    shown by blue pentagons and ETGs shown by orange pentagons.}\label{figure:mdms}
\end{figure}

We also show in Figure \ref{figure:mdms} the $M_{d}$-$M_{*}$
relationship of our simulations. We again plot our values over those
for ETGs from \citet{dariush16} and \citet{bassett17}. These
points are now compared with galaxies from the Herschel Reference
Survey \citep[HRS][]{boselli10,cortese12}, which is an infrared survey of
very nearby galaxies representing a large range in galaxy type and
environments. The proximity of this sample provides a high confidence
in visual morphological classifications as well as allowing for
measurements of $M_{d}$ to very low masses. LTGs and ETGs from HRS are
shown in Figure \ref{figure:mdms} as blue and orange pentagons,
respectively.

Similar to Figure \ref{figure:mdsfr}, our simulated galaxies are found
to occupy a similar region to dusty ETGs of \citet{dariush16} and
\citet{bassett17}, following a roughly horizontal sequence
around $M_{*}$ $\simeq$ 10$^{11}$ M$_{\odot}$. This sequence appears
to extend HRS ETGs to higher $M_{d}$ while HRS LTGs, on the other
hand, are found to follow a sequence of increasing $M_{*}$ with
increasing $M_{d}$ (albeit with relatively large scatter compared to
$M_{d}$-SFR). A relationship between $M_{*}$ and $M_{d}$ is a common
feature of star-forming galaxy samples \citep[e.g.]{cortese12b} attributed to
the fact that dust is produced by evolved stars, thus more stars will
lead to more dust. The lack of correlation seen by \citet{dariush16}
and \citet{bassett17} dusty ETGs is possible further evidence
for a merger driven origin for their dust content. In such a scenario,
the amount of dust in these galaxies appears will be dictated by the
dust content of the companion galaxy rather than the stellar mass of
the primary. Our simulated values agree well with this hypothesis as
we find nearly an order of magnitude variation in $M_{d}$ in an extremely
narrow range of $M_{*}$.  

\section{Discussion}\label{section:discussion}

\subsection{The Emergence of Counter-Rotation}

A number of previous works have investigated the emergence of
gas-versus-stellar
counter-rotation in galaxy merger simulations. The general result of
these simulations is that the primary galaxy must rotate in the
opposite direction of the merger orbit, i.e. a retrograde
merger
\citep{kormendy84,balcells90,hernquist91,bendo00,dimatteo07}. This has
also been shown by \citet{thakar97} for gas-free, sticky-particle,
N-body simulations of minor mergers. In this paper we have found a
similar result using state-of-the-art SPH simulations including
realistic stellar and gas physics, as well as tracking the production
(and destruction) of new stars, dust, and metals. 

In addition to confirming the necessity of a retrograde
orbit in producing a counter-rotating gas disk, we have also explored the
effects of including significant amounts of gas in the primary
galaxy. In such a merger, the accreted gas collides with the gas
initially in the primary galaxy, which can have a significant impact
on the final gas kinematics. We compare this with the results of
\citet{lovelace96}, who performed simulations
with smooth, filamentary accretion of gas onto a gas-rich primary
galaxy. Similar to \citet{thakar97}, smooth accretion such as this
onto a gas free primary galaxy will also result in a counter-rotating
gas disk. By gradually adding gas to the central galaxy,
\citet{lovelace96} found that once the gas content of the primary was
equal to the mass of gas accreted, the final gas disk will co-rotate
with the stellar component. For the first time, we have shown that
this is also true for the discrete accretion of a single, gas-rich
companion galaxy. This observation may influence
the discrepancy between the fraction of S0 galaxies exhibiting gas
vs stellar counter-rotation
\citep[20-40\%][]{bertola92,kuijken96,pizzella04,davis11} and the
fraction seen in later types \citep[< 8\% for Sa-Sbc
galaxies][]{kannappan01}. Galaxies with Sa and later morphologies are
far more likely to contain large amounts of gas, thus minor mergers,
even with extremely gas rich companions, will have little, if any, effect on
the observed kinematics post-merger. 

It should also be noted, however,
that S0 galaxies and LTGs have typically experienced significantly different
formation histories. The well established morphology density relation
\citep{dressler80} shows that S0 galaxies are more common in higher density, group
environments than LTGs, where minor mergers are more common
\citep[e.g.][]{wilman09}. This means that LTGs
have experienced fewer mergers, on average, and their gas content inhibits
the emergence of counter-rotating components as we have
shown. Thus, the morphology density relation provides an additional
reason for the large discrepancy between the fractions of S0 and LTGs exhibiting
counter-rotating gas. 

From an observational point of view, studies of gas-versus-stellar counter-rotation in
galaxies is becoming more commonplace due to the proliferation of
integral field spectroscopy \citep[e.g.][]{davis13,jin16}. These
studies identify not only those galaxies with gas and stellar
kinematic misalignemts of 180$^{\circ}$, those hosting
counter-rotating gas disks,
but also galaxies with misalignments between 0$^{\circ}$ and
180$^{\circ}$. The gas content of these galaxies, which is most likely
externally accreted, is expected to relax
into either co- or counter-rotation with the stellar content of the
primary galaxy within
$\sim$5.0 dynamical times \citep{davis16}. In this framework, the
distribution of kinematic misalignment angles should be strongly
peaked around 0$^{\circ}$ (this peak includes all normal/non-disturbed
galaxies), relatively flat between 0$^{\circ}$ and 180$^{\circ}$, and
weakly peaked at 180$^{\circ}$. 
This is roughly the result observed
for ATLAS$^{\textrm{3D}}$ galaxies by \citet{davis13}. 

\citet{davis16}
attempt to match the distribution in kinematic offsets found by
\citet{davis13} using a simple toy model.
Without significantly increasing the relaxation time above 5.0
dynamical times, they typically find the peak at 180$^{\circ}$ is much
larger than the observed distribution. We have shown here that in
cases where massive ETGs contain gas prior to an accretion even, the
emergence of a counter-rotating gas disk is suppressed, an effect not
accounted for in the work of \citet{davis16}. 
Recently it has been shown that $\sim$10-40\% of ETGs contain
significant amounts of molecular gas \citep[though the exact percentage may
depend on environment][]{combes07,serra12}
Properly accounting for this could suppress
the number of counter-rotating gas disks to match observations without
invoking long gas relaxation times. 

We note that ATLAS$^{\textrm{3D}}$
galaxies are, on average, of an earlier morphological type than
galaxies simulated in this work. It is reasonable to suggest this may
limit the appropriateness of a comparison between our study and the
works of \citet{davis13} and \citet{davis16}. First we point out that
the ATLAS$^{\textrm{3D}}$ sample includes a large number of S0
galaxies at a broad range of stellar masses, comparable to simulated
galaxies studied here. We have also shown in model m14 that increasing
the bulge fraction of the primary galaxy alters the efficiency of gas stripping of
the companion, but does not change the general result regarding co-
versus counter-rotating gas disks. Thus, given the simplicity of the
model presented in \citet{davis16}, for a large number of
ATLAS$^{\textrm{3D}}$ galaxies (at lower masses in particular) we
consider this comparison to be apropos. More generally, we expect our results
regarding the dependence of the emergence of counter-rotating gas
disks on merger orbits and primary gas fraction to hold for
all ETGs in a comparable mass range. 

Another interesting aspect of retrograde mergers in which the primary
galaxy initially contains significant amounts of gas is that, while the gas is
found to co-rotate with the bulk of the stars, we also identify clear
cases of stellar-versus-stellar counter-rotation. This is shown in
Figure \ref{figure:kine1} for model m8 where we find stars initially
belonging to the accreted companion galaxy end up in counter-rotating
orbits relative to those stars initially belonging to the primary galaxy. Similar
types of
orbital segregation has previously been shown in major merger
simulations by \citet{balcells98} and \citet{dimatteo07}. The reason
for this behavior is the collisionless nature of star-particles, which
is in stark contrast to the behavior of the accreted gas. We also find
that the newly formed stars have the most complex kinematic structure,
but this is due to the fact that these stars form from two, initially
distinct, gas reservoirs. Mergers such as this could explain examples
of galaxies with stellar-versus-stellar counter-rotation from observations such as
the S0 galaxy NGC 1366 \citep{morelli08,morelli17}. 

Identifying galaxies with stellar-stellar counter-rotation such as
these at higher redshift will be significantly more difficult,
particularly using observations from modern IFS surveys such as
SAMI. This is due to the fact that the spectral signature of accreted
stars will have a much lower flux density than the stars of the
primary galaxy. A possible alternative method to determine of S0 galaxies have
counter-rotating stellar populations would be to observe individual
globular clusters in nearby S0s, using techniques
demonstrated by the SLUGGs Survey \citep{pota13}. Multiple kinematic
populations of globular clusters could be evidence of galaxies hosting
both globular cluster systems formed in-situ and a population
of accreted globular clusters. 

\subsection{Gas Rings and Bar Formation}

Apparent in a number of our minor merger simulations is the emergence of
gas rings. Rings in the inner regions are quite common in disk
galaxies and can often be associated with resonances and/or bars
\citep{binney87,buta96}. Although bars have been found to form
through instabilities in simulations of isolated disk galaxies
\citep[e.g.][and references therein]{berrier16,polyachenko16}, in our
isolated disk test-case we did not observe spontaneous bar
formation. This gives us confidence that bars emerging in our
simulations are primarily the result of galaxy-galaxy interaction. 
Bar formation through galaxy interactions has long been suspected, for
example \citet{thompson81} showed that a large fraction of galaxies in
the centre of the coma cluster exhibit bars. Recent works
studying simulations of galaxy interactions, flybys in particular,
have shown that tidal forces can indeed induce bar formation, even in
cases where a secular bar would not normally appear
\citep{lang14,martinezvalpuesta16}. In our simulations, inner-rings
(with radii < 10 kpc) appear to emerge primarily in cases where a bar has also formed,
suggesting that these two structures are closely related.

Outer rings in galaxies (with radii > 10 kpc), on the other hand, are difficult
to connect with secular evolutionary processes in galaxies and are
often attributed to external accretion of gas \citep[e.g. minor
mergers][]{marino09,mapelli15,gereb16}. We find outer rings in a number of our
simulations including models m3, m5, m9, m12, and m14 (see Figures
\ref{figure:kine1} and \ref{figure:kine2}). The key
similarity between these three simulations is that the companion
galaxy in each case experiences more efficient tidal stripping of gas
during infall. In all three cases this is due to the prograde
rotation of the satellite relative to the merger orbit. This has been
known to enhance tidal stripping since the very early simulations of
\citet{toomre72}. For model m14, this process is further enhanced by
the presence of a massive bulge in the central galaxy, which produces
a deep and extended gravitational potential well.

A number of authors have explored the incidence of ring structures in
S0 galaxies finding, in general, these to be quite frequent, occurring
in $\sim$25-90\% of galaxies (depending on sample selection and
wavelength targeted). Observations have focused on indicators of star
formation such as UV \citep{salim12} or H$\alpha$ emission
\citep{pogge93} indirectly related to H$_{2}$ regions, HI
\citep[e.g.][]{oosterloo07}, and even stellar light
\citep{laurikainen13}. There are, though, relatively few
observational examples of directly observed molecular gas rings such
as that of NGC 4477 shown in \citet{crocker11}. This is partly due to
technological limitations, however our simulations suggest that a
majority of H$_{2}$ rings formed through mergers are compact, inner
rings reducing our ability to identify them in observations. This means that current
facilities will be able to resolve
molecular gas rings only in the most nearby S0 galaxies. Newer
interferometric facilities such as ALMA are thus the most likely to
observe molecular gas rings in galaxies in the near future. 

\section{Conclusions}\label{section:conclusions}

In this paper we have presented the morphologies and kinematics of
a series of minor merger remnants between two disk galaxies. These
mergers result in S0 remnants with a variety of kinematic signatures
including both co- and counter-rotation in accreted gas and stars as well as kinematically
decoupled cores and kinematic twists, which we have presented using 2D
projected maps. These maps share many similarities with IFS data
products, and, although we do not suffer from noise and surface
brightness limitations, they provide a useful comparison to maps of
galaxies from modern IFS surveys. The relatively large number of
simulations presented here has allowed us to perform a systematic
study of the
initial conditions responsible for the various kinematic signatures,
and we summarise our main results as follows:
\begin{itemize}
  \item The key factor necessary for producing counter-rotating gas
    and stellar populations in a given
    merger remnant is that the orbit of the merger must be
    retrograde with respect to the primary galaxy. Thus, accreted
    material is brought in counter to the rotational direction of the 
    stellar content of the primary galaxy.
  \item The relative spin of the companion does not affect the above
    result, however it does affect the final spatial distribution of
    gas. Encounters with prograde companion galaxy rotation result in
    more extended gas distributions than retrograde due to more
    efficient tidal stripping.
  \item If the primary galaxy contains as much or more gas than the companion,
    the accreted gas is swept up by the gas of the primary resulting
    in co-rotating gas and primary stars in the remnant. In this case, however, stars
    accreted from the companion remain counter-rotating due to their collisionless nature. This
    observation can help to explain the difference in the fraction of
    counter-rotating S0 galaxies (20-40\%) when compared to LTGs (<
    8\%).
\end{itemize}
From a practical standpoint, these conclusions show that, although
gas versus stellar counter-rotation is the easiest to observe, the
lack of such a signature does not immediately rule out recent, wet,
minor mergers. Examples of galaxies with co-rotating gas and stars may
host a secondary, counter-rotating stellar component. Such a component
may be difficult to identify in some observations, however, but we propose
the observation of counter-rotating planetary nebulae systems and
globular cluster systems in S0 galaxies \citep[the latter would be similar to observations performed by the
SLUGGS survey team][]{pota13} may provide evidence of these
systems. Finally, we also find three examples of prograde mergers with no
counter-rotating components that will be the hardest merger remnants
to identify. These three models do exhibit clear stellar, tidal
streams at large radii, which may be used to identify them as merger
remnants, although observing these will typically require extremely
long exposure times. 

In addition to these major conclusions, we observe a number of other
intriguing features in our galaxy merger simulations.
\begin{itemize}
  \item We find merger induced bars in a number of our simulations, in
    particular, retrograde mergers with gas free primary's (m2 and m3)
    exhibit the strongest bars (as well as secondary, nested
    bars). Including gas in the primary galaxy
    tends to suppress bar formation.
  \item We also find inner and outer rings in some simulations. Inner
    rings appear to be connected to bar driven resonances while outer
    rings are associated with tidally stripped gas that collects at
    large radii.
  \item Simulations in which remnants have large gas disks (m5 in
    particular) have a corresponding large angular momentum that seems
    to prevent HI from converting into H$_{2}$. This is associated
    with a significantly suppressed SFR and H$_{2}$ mass fraction.
  \item We compare the $M_{d}$-SFR and $M_{d}$-$M_{*}$ distributions
    of our simulated galaxies to observed samples of ``normal'' star-forming
    galaxies as well as massive ETGs containing dust. Our results
    support the connection between the dust content of these galaxies
    and merger activity proposed by \citet{dariush16} and \citet{bassett17}.
\end{itemize}

We note that our suite of
simulations still represents only a small portion of the possible
parameter space for minor galaxy mergers. Parameters that have not varied here
include mass ratio, dark matter fraction, and orbital inclination to
name a few. An obvious next step that can address this issue is to
apply a similar analysis to simulated galaxy mergers from the GalMer
database \citep{chilingarian10}. Regardless, the results presented
here are a useful step forward in our
understanding of the formation of counter-rotating stellar and gaseous
components in S0 galaxies.

\vspace{.75cm}
RB acknowledges support under the Australian Research Council's Discovery Projects funding
scheme (DP130100664). We also with to thank the anonymous referee for
helpful comments that have helped to clarify this manuscript.

%%%%%%%%%%%%%%%%%%%%%%%%%%%%%%%%%%%%%%%%%%%%%%%%%%



\bibliographystyle{mnras}
\bibliography{refs} % if your bibtex file is called example.bib

\begin{thebibliography}{}
\makeatletter
\relax
\def\mn@urlcharsother{\let\do\@makeother \do\$\do\&\do\#\do\^\do\_\do\%\do\~}
\def\mn@doi{\begingroup\mn@urlcharsother \@ifnextchar [ {\mn@doi@}
  {\mn@doi@[]}}
\def\mn@doi@[#1]#2{\def\@tempa{#1}\ifx\@tempa\@empty \href
  {http://dx.doi.org/#2} {doi:#2}\else \href {http://dx.doi.org/#2} {#1}\fi
  \endgroup}
\def\mn@eprint#1#2{\mn@eprint@#1:#2::\@nil}
\def\mn@eprint@arXiv#1{\href {http://arxiv.org/abs/#1} {{\tt arXiv:#1}}}
\def\mn@eprint@dblp#1{\href {http://dblp.uni-trier.de/rec/bibtex/#1.xml}
  {dblp:#1}}
\def\mn@eprint@#1:#2:#3:#4\@nil{\def\@tempa {#1}\def\@tempb {#2}\def\@tempc
  {#3}\ifx \@tempc \@empty \let \@tempc \@tempb \let \@tempb \@tempa \fi \ifx
  \@tempb \@empty \def\@tempb {arXiv}\fi \@ifundefined
  {mn@eprint@\@tempb}{\@tempb:\@tempc}{\expandafter \expandafter \csname
  mn@eprint@\@tempb\endcsname \expandafter{\@tempc}}}

\bibitem[\protect\citeauthoryear{{Athanassoula}, {Romero-G{\'o}mez}, {Bosma}
  \& {Masdemont}}{{Athanassoula} et~al.}{2009}]{athanassoula09}
{Athanassoula} E.,  {Romero-G{\'o}mez} M.,  {Bosma} A.,   {Masdemont} J.~J.,
  2009, \mn@doi [\mnras] {10.1111/j.1365-2966.2009.15583.x}, \href
  {http://adsabs.harvard.edu/abs/2009MNRAS.400.1706A} {400, 1706}

\bibitem[\protect\citeauthoryear{{Balcells} \& {Gonz{\'a}lez}}{{Balcells} \&
  {Gonz{\'a}lez}}{1998}]{balcells98}
{Balcells} M.,  {Gonz{\'a}lez} A.~C.,  1998, \mn@doi [\apjl] {10.1086/311617},
  \href {http://adsabs.harvard.edu/abs/1998ApJ...505L.109B} {505, L109}

\bibitem[\protect\citeauthoryear{{Balcells} \& {Quinn}}{{Balcells} \&
  {Quinn}}{1990}]{balcells90}
{Balcells} M.,  {Quinn} P.~J.,  1990, \mn@doi [\apj] {10.1086/169204}, \href
  {http://adsabs.harvard.edu/abs/1990ApJ...361..381B} {361, 381}

\bibitem[\protect\citeauthoryear{{Bassett} et~al.,}{{Bassett}
  et~al.}{2017}]{bassett17}
{Bassett} R.,  et~al., 2017, {The SAMI Galaxy Survey: Kinematics of Dusty
  Early-Type Galaxies}, submitted to MNRAS

\bibitem[\protect\citeauthoryear{{Bekki}}{{Bekki}}{1998}]{bekki98}
{Bekki} K.,  1998, \mn@doi [\apjl] {10.1086/311508}, \href
  {http://adsabs.harvard.edu/abs/1998ApJ...502L.133B} {502, L133}

\bibitem[\protect\citeauthoryear{{Bekki}}{{Bekki}}{2013}]{bekki13}
{Bekki} K.,  2013, \mn@doi [\mnras] {10.1093/mnras/stt589}, \href
  {http://adsabs.harvard.edu/abs/2013MNRAS.432.2298B} {432, 2298}

\bibitem[\protect\citeauthoryear{{Bekki} \& {Tsujimoto}}{{Bekki} \&
  {Tsujimoto}}{2014}]{bekki14}
{Bekki} K.,  {Tsujimoto} T.,  2014, \mn@doi [\mnras] {10.1093/mnras/stu1731},
  \href {http://adsabs.harvard.edu/abs/2014MNRAS.444.3879B} {444, 3879}

\bibitem[\protect\citeauthoryear{{Bekki}, {Couch}  \& {Shioya}}{{Bekki}
  et~al.}{2002}]{bekki02}
{Bekki} K.,  {Couch} W.~J.,   {Shioya} Y.,  2002, \mn@doi [\apj]
  {10.1086/342221}, \href {http://adsabs.harvard.edu/abs/2002ApJ...577..651B}
  {577, 651}

\bibitem[\protect\citeauthoryear{{Bendo} \& {Barnes}}{{Bendo} \&
  {Barnes}}{2000}]{bendo00}
{Bendo} G.~J.,  {Barnes} J.~E.,  2000, \mn@doi [\mnras]
  {10.1046/j.1365-8711.2000.03475.x}, \href
  {http://adsabs.harvard.edu/abs/2000MNRAS.316..315B} {316, 315}

\bibitem[\protect\citeauthoryear{{Berrier} \& {Sellwood}}{{Berrier} \&
  {Sellwood}}{2016}]{berrier16}
{Berrier} J.~C.,  {Sellwood} J.~A.,  2016, \mn@doi [\apj]
  {10.3847/0004-637X/831/1/65}, \href
  {http://adsabs.harvard.edu/abs/2016ApJ...831...65B} {831, 65}

\bibitem[\protect\citeauthoryear{{Bertola}, {Buson}  \& {Zeilinger}}{{Bertola}
  et~al.}{1992}]{bertola92}
{Bertola} F.,  {Buson} L.~M.,   {Zeilinger} W.~W.,  1992, \mn@doi [\apjl]
  {10.1086/186675}, \href {http://adsabs.harvard.edu/abs/1992ApJ...401L..79B}
  {401, L79}

\bibitem[\protect\citeauthoryear{{Binney} \& {Tremaine}}{{Binney} \&
  {Tremaine}}{1987}]{binney87}
{Binney} J.,  {Tremaine} S.,  1987, {Galactic dynamics}.
Princeton University Press

\bibitem[\protect\citeauthoryear{{Boselli} et~al.,}{{Boselli}
  et~al.}{2010}]{boselli10}
{Boselli} A.,  et~al., 2010, \mn@doi [\pasp] {10.1086/651535}, \href
  {http://adsabs.harvard.edu/abs/2010PASP..122..261B} {122, 261}

\bibitem[\protect\citeauthoryear{{Bruzual} \& {Charlot}}{{Bruzual} \&
  {Charlot}}{2003}]{bruzual03}
{Bruzual} G.,  {Charlot} S.,  2003, \mn@doi [\mnras]
  {10.1046/j.1365-8711.2003.06897.x}, \href
  {http://adsabs.harvard.edu/abs/2003MNRAS.344.1000B} {344, 1000}

\bibitem[\protect\citeauthoryear{{Bryant} et~al.,}{{Bryant}
  et~al.}{2015}]{bryant15}
{Bryant} J.~J.,  et~al., 2015, \mn@doi [\mnras] {10.1093/mnras/stu2635}, \href
  {http://adsabs.harvard.edu/abs/2015MNRAS.447.2857B} {447, 2857}

\bibitem[\protect\citeauthoryear{{Buta}, {Purcell}  \& {Crocker}}{{Buta}
  et~al.}{1996}]{buta96}
{Buta} R.,  {Purcell} G.~B.,   {Crocker} D.~A.,  1996, \mn@doi [\aj]
  {10.1086/117845}, \href {http://adsabs.harvard.edu/abs/1996AJ....111..983B}
  {111, 983}

\bibitem[\protect\citeauthoryear{{Chilingarian}, {Di Matteo}, {Combes},
  {Melchior}  \& {Semelin}}{{Chilingarian} et~al.}{2010}]{chilingarian10}
{Chilingarian} I.~V.,  {Di Matteo} P.,  {Combes} F.,  {Melchior} A.-L.,
  {Semelin} B.,  2010, \mn@doi [\aap] {10.1051/0004-6361/200912938}, \href
  {http://adsabs.harvard.edu/abs/2010A%26A...518A..61C} {518, A61}

\bibitem[\protect\citeauthoryear{{Combes}, {Young}  \& {Bureau}}{{Combes}
  et~al.}{2007}]{combes07}
{Combes} F.,  {Young} L.~M.,   {Bureau} M.,  2007, \mn@doi [\mnras]
  {10.1111/j.1365-2966.2007.11759.x}, \href
  {http://adsabs.harvard.edu/abs/2007MNRAS.377.1795C} {377, 1795}

\bibitem[\protect\citeauthoryear{{Corsini}}{{Corsini}}{2014}]{corsini14}
{Corsini} E.~M.,  2014, in {Iodice} E.,  {Corsini} E.~M.,  eds,  Astronomical
  Society of the Pacific Conference Series Vol. 486, Multi-Spin Galaxies, ASP
  Conference Series. p.~51 (\mn@eprint {arXiv} {1403.1263})

\bibitem[\protect\citeauthoryear{{Cortese} et~al.,}{{Cortese}
  et~al.}{2012a}]{cortese12b}
{Cortese} L.,  et~al., 2012a, \mn@doi [\aap] {10.1051/0004-6361/201118499},
  \href {http://adsabs.harvard.edu/abs/2012A%26A...540A..52C} {540, A52}

\bibitem[\protect\citeauthoryear{{Cortese} et~al.,}{{Cortese}
  et~al.}{2012b}]{cortese12}
{Cortese} L.,  et~al., 2012b, \mn@doi [\aap] {10.1051/0004-6361/201219312},
  \href {http://adsabs.harvard.edu/abs/2012A%26A...544A.101C} {544, A101}

\bibitem[\protect\citeauthoryear{{Crocker}, {Bureau}, {Young}  \&
  {Combes}}{{Crocker} et~al.}{2011}]{crocker11}
{Crocker} A.~F.,  {Bureau} M.,  {Young} L.~M.,   {Combes} F.,  2011, \mn@doi
  [\mnras] {10.1111/j.1365-2966.2010.17537.x}, \href
  {http://adsabs.harvard.edu/abs/2011MNRAS.410.1197C} {410, 1197}

\bibitem[\protect\citeauthoryear{{Croton} et~al.,}{{Croton}
  et~al.}{2006}]{croton06}
{Croton} D.~J.,  et~al., 2006, \mn@doi [\mnras]
  {10.1111/j.1365-2966.2005.09675.x}, \href
  {http://adsabs.harvard.edu/abs/2006MNRAS.365...11C} {365, 11}

\bibitem[\protect\citeauthoryear{{Dariush} et~al.,}{{Dariush}
  et~al.}{2016}]{dariush16}
{Dariush} A.,  et~al., 2016, \mn@doi [\mnras] {10.1093/mnras/stv2767}, \href
  {http://adsabs.harvard.edu/abs/2016MNRAS.456.2221D} {456, 2221}

\bibitem[\protect\citeauthoryear{{Davis} \& {Bureau}}{{Davis} \&
  {Bureau}}{2016}]{davis16}
{Davis} T.~A.,  {Bureau} M.,  2016, \mn@doi [\mnras] {10.1093/mnras/stv2998},
  \href {http://adsabs.harvard.edu/abs/2016MNRAS.457..272D} {457, 272}

\bibitem[\protect\citeauthoryear{{Davis}, {Efstathiou}, {Frenk}  \&
  {White}}{{Davis} et~al.}{1985}]{davis85}
{Davis} M.,  {Efstathiou} G.,  {Frenk} C.~S.,   {White} S.~D.~M.,  1985,
  \mn@doi [\apj] {10.1086/163168}, \href
  {http://adsabs.harvard.edu/abs/1985ApJ...292..371D} {292, 371}

\bibitem[\protect\citeauthoryear{{Davis} et~al.,}{{Davis}
  et~al.}{2011}]{davis11}
{Davis} T.~A.,  et~al., 2011, \mn@doi [\mnras]
  {10.1111/j.1365-2966.2011.19355.x}, \href
  {http://adsabs.harvard.edu/abs/2011MNRAS.417..882D} {417, 882}

\bibitem[\protect\citeauthoryear{{Davis} et~al.,}{{Davis}
  et~al.}{2013}]{davis13}
{Davis} T.~A.,  et~al., 2013, \mn@doi [\mnras] {10.1093/mnras/sts353}, \href
  {http://adsabs.harvard.edu/abs/2013MNRAS.429..534D} {429, 534}

\bibitem[\protect\citeauthoryear{{Deng} \& {Yu}}{{Deng} \& {Yu}}{2015}]{deng15}
{Deng} X.-F.,  {Yu} G.,  2015, \mn@doi [Astrophysics]
  {10.1007/s10511-015-9380-y}, \href
  {http://adsabs.harvard.edu/abs/2015Ap.....58..250D} {58, 250}

\bibitem[\protect\citeauthoryear{{Di Matteo}, {Combes}, {Melchior}  \&
  {Semelin}}{{Di Matteo} et~al.}{2007}]{dimatteo07}
{Di Matteo} P.,  {Combes} F.,  {Melchior} A.-L.,   {Semelin} B.,  2007, \mn@doi
  [\aap] {10.1051/0004-6361:20066959}, \href
  {http://adsabs.harvard.edu/abs/2007A%26A...468...61D} {468, 61}

\bibitem[\protect\citeauthoryear{{Draine} \& {Salpeter}}{{Draine} \&
  {Salpeter}}{1979}]{draine79}
{Draine} B.~T.,  {Salpeter} E.~E.,  1979, \mn@doi [\apj] {10.1086/157165},
  \href {http://adsabs.harvard.edu/abs/1979ApJ...231...77D} {231, 77}

\bibitem[\protect\citeauthoryear{{Dressler}}{{Dressler}}{1980}]{dressler80}
{Dressler} A.,  1980, \mn@doi [\apj] {10.1086/157753}, \href
  {http://adsabs.harvard.edu/abs/1980ApJ...236..351D} {236, 351}

\bibitem[\protect\citeauthoryear{{Erwin}}{{Erwin}}{2002}]{erwin02}
{Erwin} P.,  2002, in {Athanassoula} E.,  {Bosma} A.,   {Mujica} R.,  eds,
  Astronomical Society of the Pacific Conference Series Vol. 275, Disks of
  Galaxies: Kinematics, Dynamics and Peturbations. pp 271--274

\bibitem[\protect\citeauthoryear{{Erwin}}{{Erwin}}{2004}]{erwin04}
{Erwin} P.,  2004, \mn@doi [\aap] {10.1051/0004-6361:20034408}, \href
  {http://adsabs.harvard.edu/abs/2004A%26A...415..941E} {415, 941}

\bibitem[\protect\citeauthoryear{{Galletta}}{{Galletta}}{1987}]{galletta87}
{Galletta} G.,  1987, \mn@doi [\apj] {10.1086/165389}, \href
  {http://adsabs.harvard.edu/abs/1987ApJ...318..531G} {318, 531}

\bibitem[\protect\citeauthoryear{{Gereb}, {Catinella}, {Cortese}, {Bekki},
  {Moran}  \& {Schiminovich}}{{Gereb} et~al.}{2016}]{gereb16}
{Gereb} K.,  {Catinella} B.,  {Cortese} L.,  {Bekki} K.,  {Moran} S.,
  {Schiminovich} D.,  2016, preprint, \href
  {http://adsabs.harvard.edu/abs/2016arXiv160701446G} {} (\mn@eprint {arXiv}
  {1607.01446})

\bibitem[\protect\citeauthoryear{{Gunn} \& {Gott}}{{Gunn} \&
  {Gott}}{1972}]{gunn72}
{Gunn} J.~E.,  {Gott} III J.~R.,  1972, \mn@doi [\apj] {10.1086/151605}, \href
  {http://adsabs.harvard.edu/abs/1972ApJ...176....1G} {176, 1}

\bibitem[\protect\citeauthoryear{{Hernquist} \& {Barnes}}{{Hernquist} \&
  {Barnes}}{1991}]{hernquist91}
{Hernquist} L.,  {Barnes} J.~E.,  1991, \mn@doi [\nat] {10.1038/354210a0},
  \href {http://adsabs.harvard.edu/abs/1991Natur.354..210H} {354, 210}

\bibitem[\protect\citeauthoryear{{Huang}, {Haynes}, {Giovanelli}, {Brinchmann},
  {Stierwalt}  \& {Neff}}{{Huang} et~al.}{2012}]{huang12}
{Huang} S.,  {Haynes} M.~P.,  {Giovanelli} R.,  {Brinchmann} J.,  {Stierwalt}
  S.,   {Neff} S.~G.,  2012, \mn@doi [\aj] {10.1088/0004-6256/143/6/133}, \href
  {http://adsabs.harvard.edu/abs/2012AJ....143..133H} {143, 133}

\bibitem[\protect\citeauthoryear{{Jesseit}, {Naab}, {Peletier}  \&
  {Burkert}}{{Jesseit} et~al.}{2007}]{jesseit07}
{Jesseit} R.,  {Naab} T.,  {Peletier} R.~F.,   {Burkert} A.,  2007, \mn@doi
  [\mnras] {10.1111/j.1365-2966.2007.11524.x}, \href
  {http://adsabs.harvard.edu/abs/2007MNRAS.376..997J} {376, 997}

\bibitem[\protect\citeauthoryear{{Jin} et~al.,}{{Jin} et~al.}{2016}]{jin16}
{Jin} Y.,  et~al., 2016, \mn@doi [\mnras] {10.1093/mnras/stw2055}, \href
  {http://adsabs.harvard.edu/abs/2016MNRAS.463..913J} {463, 913}

\bibitem[\protect\citeauthoryear{{Kannappan} \& {Fabricant}}{{Kannappan} \&
  {Fabricant}}{2001}]{kannappan01}
{Kannappan} S.~J.,  {Fabricant} D.~G.,  2001, \mn@doi [\aj] {10.1086/318027},
  \href {http://adsabs.harvard.edu/abs/2001AJ....121..140K} {121, 140}

\bibitem[\protect\citeauthoryear{{Kennicutt}}{{Kennicutt}}{1998}]{kenni98}
{Kennicutt} Jr. R.~C.,  1998, \mn@doi [\apj] {10.1086/305588}, \href
  {http://adsabs.harvard.edu/abs/1998ApJ...498..541K} {498, 541}

\bibitem[\protect\citeauthoryear{{Kormendy}}{{Kormendy}}{1984}]{kormendy84}
{Kormendy} J.,  1984, \mn@doi [\apj] {10.1086/162717}, \href
  {http://adsabs.harvard.edu/abs/1984ApJ...287..577K} {287, 577}

\bibitem[\protect\citeauthoryear{{Kormendy} \& {Bender}}{{Kormendy} \&
  {Bender}}{2012}]{kormendy12}
{Kormendy} J.,  {Bender} R.,  2012, \mn@doi [\apjs]
  {10.1088/0067-0049/198/1/2}, \href
  {http://adsabs.harvard.edu/abs/2012ApJS..198....2K} {198, 2}

\bibitem[\protect\citeauthoryear{{Kroupa}}{{Kroupa}}{2001}]{kroupa01}
{Kroupa} P.,  2001, \mn@doi [\mnras] {10.1046/j.1365-8711.2001.04022.x}, \href
  {http://adsabs.harvard.edu/abs/2001MNRAS.322..231K} {322, 231}

\bibitem[\protect\citeauthoryear{{Kuijken}, {Fisher}  \&
  {Merrifield}}{{Kuijken} et~al.}{1996}]{kuijken96}
{Kuijken} K.,  {Fisher} D.,   {Merrifield} M.~R.,  1996, \mn@doi [\mnras]
  {10.1093/mnras/283.2.543}, \href
  {http://adsabs.harvard.edu/abs/1996MNRAS.283..543K} {283, 543}

\bibitem[\protect\citeauthoryear{{Lang}, {Holley-Bockelmann}  \&
  {Sinha}}{{Lang} et~al.}{2014}]{lang14}
{Lang} M.,  {Holley-Bockelmann} K.,   {Sinha} M.,  2014, \mn@doi [\apjl]
  {10.1088/2041-8205/790/2/L33}, \href
  {http://adsabs.harvard.edu/abs/2014ApJ...790L..33L} {790, L33}

\bibitem[\protect\citeauthoryear{{Laurikainen}, {Salo}, {Athanassoula},
  {Bosma}, {Buta}  \& {Janz}}{{Laurikainen} et~al.}{2013}]{laurikainen13}
{Laurikainen} E.,  {Salo} H.,  {Athanassoula} E.,  {Bosma} A.,  {Buta} R.,
  {Janz} J.,  2013, \mn@doi [\mnras] {10.1093/mnras/stt150}, \href
  {http://adsabs.harvard.edu/abs/2013MNRAS.430.3489L} {430, 3489}

\bibitem[\protect\citeauthoryear{{Le Borgne}, {Rocca-Volmerange}, {Prugniel},
  {Lan{\c c}on}, {Fioc}  \& {Soubiran}}{{Le Borgne} et~al.}{2004}]{leborg04}
{Le Borgne} D.,  {Rocca-Volmerange} B.,  {Prugniel} P.,  {Lan{\c c}on} A.,
  {Fioc} M.,   {Soubiran} C.,  2004, \mn@doi [\aap]
  {10.1051/0004-6361:200400044}, \href
  {http://adsabs.harvard.edu/abs/2004A%26A...425..881L} {425, 881}

\bibitem[\protect\citeauthoryear{{Lotz}, {Jonsson}, {Cox}, {Croton}, {Primack},
  {Somerville}  \& {Stewart}}{{Lotz} et~al.}{2011}]{lotz11}
{Lotz} J.~M.,  {Jonsson} P.,  {Cox} T.~J.,  {Croton} D.,  {Primack} J.~R.,
  {Somerville} R.~S.,   {Stewart} K.,  2011, \mn@doi [\apj]
  {10.1088/0004-637X/742/2/103}, \href
  {http://adsabs.harvard.edu/abs/2011ApJ...742..103L} {742, 103}

\bibitem[\protect\citeauthoryear{{Lovelace} \& {Chou}}{{Lovelace} \&
  {Chou}}{1996}]{lovelace96}
{Lovelace} R.~V.~E.,  {Chou} T.,  1996, \mn@doi [\apjl] {10.1086/310232}, \href
  {http://adsabs.harvard.edu/abs/1996ApJ...468L..25L} {468, L25}

\bibitem[\protect\citeauthoryear{{Lynden-Bell}}{{Lynden-Bell}}{1960}]{lyndenbell60}
{Lynden-Bell} D.,  1960, \mn@doi [\mnras] {10.1093/mnras/120.3.204}, \href
  {http://adsabs.harvard.edu/abs/1960MNRAS.120..204L} {120, 204}

\bibitem[\protect\citeauthoryear{{Mapelli}}{{Mapelli}}{2015}]{mapelli15}
{Mapelli} M.,  2015, \mn@doi [Galaxies] {10.3390/galaxies3040192}, \href
  {http://adsabs.harvard.edu/abs/2015Galax...3..192M} {3, 192}

\bibitem[\protect\citeauthoryear{{Marino} et~al.,}{{Marino}
  et~al.}{2009}]{marino09}
{Marino} A.,  et~al., 2009, \mn@doi [\aap] {10.1051/0004-6361/200911819}, \href
  {http://adsabs.harvard.edu/abs/2009A%26A...508.1235M} {508, 1235}

\bibitem[\protect\citeauthoryear{{Marinova} et~al.,}{{Marinova}
  et~al.}{2012}]{marinova12}
{Marinova} I.,  et~al., 2012, \mn@doi [\apj] {10.1088/0004-637X/746/2/136},
  \href {http://adsabs.harvard.edu/abs/2012ApJ...746..136M} {746, 136}

\bibitem[\protect\citeauthoryear{{Martinez-Valpuesta}, {Aguerri}, {C{\'e}sar
  Gonz{\'a}lez-Garc{\'{\i}}a}, {Dalla Vecchia}  \&
  {Stringer}}{{Martinez-Valpuesta} et~al.}{2016}]{martinezvalpuesta16}
{Martinez-Valpuesta} I.,  {Aguerri} J.~A.~L.,  {C{\'e}sar
  Gonz{\'a}lez-Garc{\'{\i}}a} A.,  {Dalla Vecchia} C.,   {Stringer} M.,  2016,
  preprint, \href {http://adsabs.harvard.edu/abs/2016arXiv161002326M} {}
  (\mn@eprint {arXiv} {1610.02326})

\bibitem[\protect\citeauthoryear{{Mihos}, {Walker}, {Hernquist}, {Mendes de
  Oliveira}  \& {Bolte}}{{Mihos} et~al.}{1995}]{mihos95}
{Mihos} J.~C.,  {Walker} I.~R.,  {Hernquist} L.,  {Mendes de Oliveira} C.,
  {Bolte} M.,  1995, \mn@doi [\apjl] {10.1086/309576}, \href
  {http://adsabs.harvard.edu/abs/1995ApJ...447L..87M} {447, L87}

\bibitem[\protect\citeauthoryear{{Moffett} et~al.,}{{Moffett}
  et~al.}{2016}]{moffett16}
{Moffett} A.~J.,  et~al., 2016, \mn@doi [\mnras] {10.1093/mnras/stv2883}, \href
  {http://adsabs.harvard.edu/abs/2016MNRAS.457.1308M} {457, 1308}

\bibitem[\protect\citeauthoryear{{Morelli} et~al.,}{{Morelli}
  et~al.}{2008}]{morelli08}
{Morelli} L.,  et~al., 2008, \mn@doi [\mnras]
  {10.1111/j.1365-2966.2008.13566.x}, \href
  {http://adsabs.harvard.edu/abs/2008MNRAS.389..341M} {389, 341}

\bibitem[\protect\citeauthoryear{{Morelli} et~al.,}{{Morelli}
  et~al.}{2017}]{morelli17}
{Morelli} L.,  et~al., 2017, preprint, \href
  {http://adsabs.harvard.edu/abs/2017arXiv170107631M} {} (\mn@eprint {arXiv}
  {1701.07631})

\bibitem[\protect\citeauthoryear{{Nakamura}, {Fukugita}, {Yasuda}, {Loveday},
  {Brinkmann}, {Schneider}, {Shimasaku}  \& {SubbaRao}}{{Nakamura}
  et~al.}{2003}]{nakamura03}
{Nakamura} O.,  {Fukugita} M.,  {Yasuda} N.,  {Loveday} J.,  {Brinkmann} J.,
  {Schneider} D.~P.,  {Shimasaku} K.,   {SubbaRao} M.,  2003, \mn@doi [\aj]
  {10.1086/368135}, \href {http://adsabs.harvard.edu/abs/2003AJ....125.1682N}
  {125, 1682}

\bibitem[\protect\citeauthoryear{{Namekata}, {Habe}, {Matsui}  \&
  {Saitoh}}{{Namekata} et~al.}{2009}]{namekata09}
{Namekata} D.,  {Habe} A.,  {Matsui} H.,   {Saitoh} T.~R.,  2009, \mn@doi
  [\apj] {10.1088/0004-637X/691/2/1525}, \href
  {http://adsabs.harvard.edu/abs/2009ApJ...691.1525N} {691, 1525}

\bibitem[\protect\citeauthoryear{{Navarro}, {Frenk}  \& {White}}{{Navarro}
  et~al.}{1996}]{navarro96}
{Navarro} J.~F.,  {Frenk} C.~S.,   {White} S.~D.~M.,  1996, \mn@doi [\apj]
  {10.1086/177173}, \href {http://adsabs.harvard.edu/abs/1996ApJ...462..563N}
  {462, 563}

\bibitem[\protect\citeauthoryear{{Neto} et~al.,}{{Neto} et~al.}{2007}]{neto07}
{Neto} A.~F.,  et~al., 2007, \mn@doi [\mnras]
  {10.1111/j.1365-2966.2007.12381.x}, \href
  {http://adsabs.harvard.edu/abs/2007MNRAS.381.1450N} {381, 1450}

\bibitem[\protect\citeauthoryear{{Noordermeer}}{{Noordermeer}}{2008}]{noordermeer08}
{Noordermeer} E.,  2008, \mn@doi [\mnras] {10.1111/j.1365-2966.2008.12837.x},
  \href {http://adsabs.harvard.edu/abs/2008MNRAS.385.1359N} {385, 1359}

\bibitem[\protect\citeauthoryear{{Oosterloo}, {Morganti}, {Sadler}, {van der
  Hulst}  \& {Serra}}{{Oosterloo} et~al.}{2007}]{oosterloo07}
{Oosterloo} T.~A.,  {Morganti} R.,  {Sadler} E.~M.,  {van der Hulst} T.,
  {Serra} P.,  2007, \mn@doi [\aap] {10.1051/0004-6361:20066384}, \href
  {http://adsabs.harvard.edu/abs/2007A%26A...465..787O} {465, 787}

\bibitem[\protect\citeauthoryear{{Pizzella}, {Corsini}, {Vega Beltr{\'a}n}  \&
  {Bertola}}{{Pizzella} et~al.}{2004}]{pizzella04}
{Pizzella} A.,  {Corsini} E.~M.,  {Vega Beltr{\'a}n} J.~C.,   {Bertola} F.,
  2004, \mn@doi [\aap] {10.1051/0004-6361:20047183}, \href
  {http://adsabs.harvard.edu/abs/2004A%26A...424..447P} {424, 447}

\bibitem[\protect\citeauthoryear{{Pogge} \& {Eskridge}}{{Pogge} \&
  {Eskridge}}{1993}]{pogge93}
{Pogge} R.~W.,  {Eskridge} P.~B.,  1993, \mn@doi [\aj] {10.1086/116735}, \href
  {http://adsabs.harvard.edu/abs/1993AJ....106.1405P} {106, 1405}

\bibitem[\protect\citeauthoryear{{Polyachenko}, {Berczik}  \&
  {Just}}{{Polyachenko} et~al.}{2016}]{polyachenko16}
{Polyachenko} E.~V.,  {Berczik} P.,   {Just} A.,  2016, \mn@doi [\mnras]
  {10.1093/mnras/stw1907}, \href
  {http://adsabs.harvard.edu/abs/2016MNRAS.462.3727P} {462, 3727}

\bibitem[\protect\citeauthoryear{{Pota} et~al.,}{{Pota} et~al.}{2013}]{pota13}
{Pota} V.,  et~al., 2013, \mn@doi [\mnras] {10.1093/mnras/sts029}, \href
  {http://adsabs.harvard.edu/abs/2013MNRAS.428..389P} {428, 389}

\bibitem[\protect\citeauthoryear{{Press} \& {Schechter}}{{Press} \&
  {Schechter}}{1974}]{press74}
{Press} W.~H.,  {Schechter} P.,  1974, \mn@doi [\apj] {10.1086/152650}, \href
  {http://adsabs.harvard.edu/abs/1974ApJ...187..425P} {187, 425}

\bibitem[\protect\citeauthoryear{{Querejeta} et~al.,}{{Querejeta}
  et~al.}{2015}]{querejeta15}
{Querejeta} M.,  et~al., 2015, \mn@doi [\aap] {10.1051/0004-6361/201526354},
  \href {http://adsabs.harvard.edu/abs/2015A%26A...579L...2Q} {579, L2}

\bibitem[\protect\citeauthoryear{{Rubin}}{{Rubin}}{1994}]{rubin94}
{Rubin} V.~C.,  1994, \mn@doi [\aj] {10.1086/117083}, \href
  {http://adsabs.harvard.edu/abs/1994AJ....108..456R} {108, 456}

\bibitem[\protect\citeauthoryear{{Rubin}, {Graham}  \& {Kenney}}{{Rubin}
  et~al.}{1992}]{rubin92}
{Rubin} V.~C.,  {Graham} J.~A.,   {Kenney} J.~D.~P.,  1992, \mn@doi [\apjl]
  {10.1086/186460}, \href {http://adsabs.harvard.edu/abs/1992ApJ...394L...9R}
  {394, L9}

\bibitem[\protect\citeauthoryear{{Salim}, {Fang}, {Rich}, {Faber}  \&
  {Thilker}}{{Salim} et~al.}{2012}]{salim12}
{Salim} S.,  {Fang} J.~J.,  {Rich} R.~M.,  {Faber} S.~M.,   {Thilker} D.~A.,
  2012, \mn@doi [\apj] {10.1088/0004-637X/755/2/105}, \href
  {http://adsabs.harvard.edu/abs/2012ApJ...755..105S} {755, 105}

\bibitem[\protect\citeauthoryear{{Sandage}}{{Sandage}}{2004}]{sandage04}
{Sandage} A.,  2004, in {Block} D.~L.,  {Puerari} I.,  {Freeman} K.~C.,
  {Groess} R.,   {Block} E.~K.,  eds,  Astrophysics and Space Science Library
  Vol. 319, Penetrating Bars Through Masks of Cosmic Dust. p.~39,
  \mn@doi{10.1007/978-1-4020-2862-5_3}

\bibitem[\protect\citeauthoryear{{Schmidt}}{{Schmidt}}{1959}]{schmidt59}
{Schmidt} M.,  1959, \mn@doi [\apj] {10.1086/146614}, \href
  {http://adsabs.harvard.edu/abs/1959ApJ...129..243S} {129, 243}

\bibitem[\protect\citeauthoryear{{Serra} et~al.,}{{Serra}
  et~al.}{2012}]{serra12}
{Serra} P.,  et~al., 2012, \mn@doi [\mnras] {10.1111/j.1365-2966.2012.20219.x},
  \href {http://adsabs.harvard.edu/abs/2012MNRAS.422.1835S} {422, 1835}

\bibitem[\protect\citeauthoryear{{Shlosman}, {Frank}  \& {Begelman}}{{Shlosman}
  et~al.}{1989}]{shlosman89}
{Shlosman} I.,  {Frank} J.,   {Begelman} M.~C.,  1989, \mn@doi [\nat]
  {10.1038/338045a0}, \href {http://adsabs.harvard.edu/abs/1989Natur.338...45S}
  {338, 45}

\bibitem[\protect\citeauthoryear{{Somerville}, {Hopkins}, {Cox}, {Robertson}
  \& {Hernquist}}{{Somerville} et~al.}{2008}]{somerville08}
{Somerville} R.~S.,  {Hopkins} P.~F.,  {Cox} T.~J.,  {Robertson} B.~E.,
  {Hernquist} L.,  2008, \mn@doi [\mnras] {10.1111/j.1365-2966.2008.13805.x},
  \href {http://adsabs.harvard.edu/abs/2008MNRAS.391..481S} {391, 481}

\bibitem[\protect\citeauthoryear{{Stewart}, {Bullock}, {Wechsler}, {Maller}  \&
  {Zentner}}{{Stewart} et~al.}{2008}]{stewart08}
{Stewart} K.~R.,  {Bullock} J.~S.,  {Wechsler} R.~H.,  {Maller} A.~H.,
  {Zentner} A.~R.,  2008, \mn@doi [\apj] {10.1086/588579}, \href
  {http://adsabs.harvard.edu/abs/2008ApJ...683..597S} {683, 597}

\bibitem[\protect\citeauthoryear{{Thakar}, {Ryden}, {Jore}  \&
  {Broeils}}{{Thakar} et~al.}{1997}]{thakar97}
{Thakar} A.~R.,  {Ryden} B.~S.,  {Jore} K.~P.,   {Broeils} A.~H.,  1997, \apj,
  \href {http://adsabs.harvard.edu/abs/1997ApJ...479..702T} {479, 702}

\bibitem[\protect\citeauthoryear{{Thompson}}{{Thompson}}{1981}]{thompson81}
{Thompson} L.~A.,  1981, \mn@doi [\apjl] {10.1086/183476}, \href
  {http://adsabs.harvard.edu/abs/1981ApJ...244L..43T} {244, L43}

\bibitem[\protect\citeauthoryear{{Toomre}}{{Toomre}}{1982}]{toomre82}
{Toomre} A.,  1982, \mn@doi [\apj] {10.1086/160190}, \href
  {http://adsabs.harvard.edu/abs/1982ApJ...259..535T} {259, 535}

\bibitem[\protect\citeauthoryear{{Toomre} \& {Toomre}}{{Toomre} \&
  {Toomre}}{1972}]{toomre72}
{Toomre} A.,  {Toomre} J.,  1972, in Bulletin of the American Astronomical
  Society. p.~214

\bibitem[\protect\citeauthoryear{{Tremonti} et~al.,}{{Tremonti}
  et~al.}{2004}]{tremonti04}
{Tremonti} C.~A.,  et~al., 2004, \mn@doi [\apj] {10.1086/423264}, \href
  {http://adsabs.harvard.edu/abs/2004ApJ...613..898T} {613, 898}

\bibitem[\protect\citeauthoryear{{Tsujimoto}, {Nomoto}, {Yoshii}, {Hashimoto},
  {Yanagida}  \& {Thielemann}}{{Tsujimoto} et~al.}{1995}]{tsujimoto95}
{Tsujimoto} T.,  {Nomoto} K.,  {Yoshii} Y.,  {Hashimoto} M.,  {Yanagida} S.,
  {Thielemann} F.-K.,  1995, \mn@doi [\mnras] {10.1093/mnras/277.3.945}, \href
  {http://adsabs.harvard.edu/abs/1995MNRAS.277..945T} {277, 945}

\bibitem[\protect\citeauthoryear{{Whitaker}, {van Dokkum}, {Brammer}  \&
  {Franx}}{{Whitaker} et~al.}{2012}]{whitaker12}
{Whitaker} K.~E.,  {van Dokkum} P.~G.,  {Brammer} G.,   {Franx} M.,  2012,
  \mn@doi [\apjl] {10.1088/2041-8205/754/2/L29}, \href
  {http://adsabs.harvard.edu/abs/2012ApJ...754L..29W} {754, L29}

\bibitem[\protect\citeauthoryear{{White} \& {Rees}}{{White} \&
  {Rees}}{1978}]{white78}
{White} S.~D.~M.,  {Rees} M.~J.,  1978, \mnras, \href
  {http://adsabs.harvard.edu/abs/1978MNRAS.183..341W} {183, 341}

\bibitem[\protect\citeauthoryear{{Wilman}, {Oemler}, {Mulchaey}, {McGee},
  {Balogh}  \& {Bower}}{{Wilman} et~al.}{2009}]{wilman09}
{Wilman} D.~J.,  {Oemler} Jr. A.,  {Mulchaey} J.~S.,  {McGee} S.~L.,  {Balogh}
  M.~L.,   {Bower} R.~G.,  2009, \mn@doi [\apj] {10.1088/0004-637X/692/1/298},
  \href {http://adsabs.harvard.edu/abs/2009ApJ...692..298W} {692, 298}

\bibitem[\protect\citeauthoryear{{Wozniak}}{{Wozniak}}{2015}]{wozniak15}
{Wozniak} H.,  2015, \mn@doi [\aap] {10.1051/0004-6361/201425005}, \href
  {http://adsabs.harvard.edu/abs/2015A%26A...575A...7W} {575, A7}

\bibitem[\protect\citeauthoryear{{Wozniak}, {Friedli}, {Martinet}, {Martin}  \&
  {Bratschi}}{{Wozniak} et~al.}{1995}]{wozniak95}
{Wozniak} H.,  {Friedli} D.,  {Martinet} L.,  {Martin} P.,   {Bratschi} P.,
  1995, \aaps, \href {http://adsabs.harvard.edu/abs/1995A%26AS..111..115W}
  {111, 115}

\bibitem[\protect\citeauthoryear{{da Cunha}, {Eminian}, {Charlot}  \&
  {Blaizot}}{{da Cunha} et~al.}{2010}]{dacunha10a}
{da Cunha} E.,  {Eminian} C.,  {Charlot} S.,   {Blaizot} J.,  2010, \mn@doi
  [\mnras] {10.1111/j.1365-2966.2010.16344.x}, \href
  {http://adsabs.harvard.edu/abs/2010MNRAS.403.1894D} {403, 1894}

\bibitem[\protect\citeauthoryear{{van den Hoek} \& {Groenewegen}}{{van den
  Hoek} \& {Groenewegen}}{1997}]{vandenhoek97}
{van den Hoek} L.~B.,  {Groenewegen} M.~A.~T.,  1997, \mn@doi [\aaps]
  {10.1051/aas:1997162}, \href
  {http://adsabs.harvard.edu/abs/1997A%26AS..123..305V} {123}

\makeatother
\end{thebibliography}

%%%%%%%%%%%%%%%%%%%%%%%%%%%%%%%%%%%%%%%%%%%%%%%%%%

%%%%%%%%%%%%%%%%% APPENDICES %%%%%%%%%%%%%%%%%%%%%

\appendix
\newpage
\section{Table of Integrated Values}

Here we present Table \ref{table:3}, which gives the integrated
quantities calculated as described in Section
\ref{section:intquant}. These include change in gas mass, final
half mass radii and concentrations of each component, and the change
in angular momentum of gas and primar stars.  

%%%%% TABLE3
\begin{table*}\label{table:3}
\centering
\begin{minipage}{\textwidth}
\begin{center}
\caption{Integrated Quantities}
\begin{tabular}{|l|c|c|c|c|c|c|c|c|c|c|c|}
\hline
{ ID \footnote{ The isolated model is m1. 
 }} & 
{ $\Delta M_{g}$ \footnote{ percentage of gas mass converted into
      stars
 }} & 
{ $\Delta L_{g}$ \footnote{ percentage of angular momentum lost by
      gas component, note in models with non-zero $f_{g,p}$ this can
      often be negative (net gain of $L$) as gas converted to
      stars represents a loss of $L$ typically occuring in low $L$ gas.
 }} & 
{ $r_{f50,g}$ \footnote{ Final half-mass radius of gas component in kpc
 }} &
{ $c_{g}$ \footnote{ Final concentration index, gas
 }} & 
{ $\Delta L_{p,*}$ \footnote{ percentage of angular momentum lost
      by primary stars
 }} & 
{ $r_{f50,p*}$  \footnote{ Final half-mass radius of primary stars in kpc
}} &
{ $c_{p*}$ \footnote{ Final concentration index, primary stars
 }} &
{ $r_{f50,n*}$  \footnote{ Final half-mass radius of new stars
     in kpc
}} &
{ $c_{n*}$ \footnote{ Final concentration index, new stars
 }} &
{ $r_{f50,c*}$  \footnote{ Final half-mass radius of companion stars in kpc
}} &
{ $c_{c*}$ \footnote{ Final concentration index, companion stars
 }} \\
\hline
\hline
m1 & 5.8\% & -4.8\% & 5.9 & 2.36 & -47.2\% & 0.9 & 2.43 & n/a & n/a &
                                                                      n/a
   & n/a\\
m2 & 60.7\% & 84.7\% & 3.1 & 2.20 & 16.9\% & 3.6 & 3.76 & 3.4 & 4.33 &
                                                                       14.6
   & 1.63\\
m3 & 40.8\% & 47.9\% & 13.1 & 2.14 & 13.9\% & 3.9 & 3.58 & 2.9 & 6.04
    & 16.6 & 1.68\\
m4 & 44.9\% & 51.4\% & 10.4 & 1.46 & 10.1\% & 4.1 & 3.79 & 6.6 & 1.83
    & 9.9 & 1.78\\
m5 & 57.6\% & 12.3\% & 21.6 & 1.50 & 8.0\% & 4.1 & 3.91 & 0.6 & 20.2 &
                                                                      12.9
   &  1.91\\
m6 & 69.7\% & 65.2\% & 6.1 & 1.53 & 2.7\% & 4.6 & 2.57 & 1.4 & 4.64 &
                                                                      12.1
   & 1.76 \\
m7 & 91.8\% & -6.2\% & 7.6 & 1.59 & 3.2\% & 4.9 & 2.59 & 0.9 & 3.29 &
                                                                      12.9
   & 1.62 \\
m8 & 80.4\% & -173.1\% & 10.6 & 1.59 & -1.1\% & 4.9 & 2.64 & 0.9 &
                                                                   7.29
    & 12.1 & 1.85 \\
m9 & 77.1\% & -93.1\% & 13.6 & 2.32 & -3.8\% & 4.9 & 2.59 & 1.1 & 5.22
    & 15.6 & 1.75 \\
m10 & 28.4\% & 70.1\% & 10.9 & 1.25 & 7.0\% & 4.4 & 3.00 & 11.6 & 2.37
    & 19.4 & 1.65 \\
m11 & 46.9\% & 68.8\% & 9.1 & 1.27 & 3.9\% & 4.6 & 2.73 & 7.9 & 2.56 &
                                                                       19.6
   & 1.70 \\
m12 & 62.7\% & -41.5\% & 16.9 & 1.68 & -2.4\% & 4.4 & 3.69 & 1.4 &
                                                                   5.73
    & 13.4 & 1.90 \\
m13 & 22.1\% & 45.9\% & 7.9 & x & -2.4\% & 4.4 & 3.11 & 56.1 & 1.99 &
                                                                      56.1
   & 1.86 \\
m14 & 34.3\% & 36.4\% & 21.1 & 2.13 & 0.3\% & 3.1 & 3.48 & 1.6 & 22.23
    & 28.4 & 2.09\\
m15 & 15.2\% & 81.4\% & 5.4 & 1.37 & 6.6\% & 4.4 & 2.94 & 8.4 & 1.81 &
                                                                       17.4
   & 1.58 \\
m16 & 8.4\% & 75.5\% & 6.4 & 1.35 & 5.6\% & 4.6 & 2.73 & 8.1 & 1.77 &
                                                                      17.4
   & 1.56 \\
\hline
\end{tabular}
\end{center}
\end{minipage}
\end{table*}

%%%%%%%%%%%%%%%%%%%%%%%%%%%%%%%%%%%%%%%%%%%%%%%%%%

% Don't change these lines
\bsp	% typesetting comment
\label{lastpage}
\end{document}